\documentclass[amssymb,amsmath,aps,showpacs,floatfix,nofootinbib,showpacs,12pt]{revtex4}
\usepackage{amssymb}
\usepackage{graphicx}
\usepackage{color}
\usepackage{soul}
\usepackage{latexsym}

\begin{document}

\voffset 1.25cm

\title{ Clumpiness enhancement of charged cosmic rays from dark matter 
annihilation with Sommerfeld effect}

\author{Qiang Yuan$^{1}$, Xiao-Jun Bi$^{2,1}$, Jia Liu$^{3}$, 
Peng-Fei Yin$^{3}$, Juan Zhang$^{1}$ and Shou-Hua Zhu$^{3}$}

\affiliation{$^{1}$ Key Laboratory of Particle Astrophysics,
Institute of High Energy Physics, Chinese
Academy of Sciences, Beijing 100049, P. R. China \\
$^{2}$ Center for High Energy Physics,
Peking University, Beijing 100871, P.R. China \\
$^{3}$ Institute of Theoretical Physics \& State Key Laboratory of
Nuclear Physics and Technology, Peking University, Beijing 100871,
P.R. China  }

\date{\today}

\begin{abstract}

Boost factors of dark matter annihilation into antiprotons and
electrons/positrons due to the clumpiness of dark matter distribution
are studied in detail in this work, taking the Sommerfeld effect
into account. It has been thought that the Sommerfeld effect, if
exists, will be more remarkable in substructures because they are
colder than the host halo, and may result in a larger boost factor.
We give a full calculation of the boost factors based on the recent
N-body simulations. Three typical cases of Sommerfeld effects, the
non-resonant, moderately resonant and strongly resonant cases are
considered. We find that for the non-resonant and moderately
resonant cases the enhancement effects of substructures due to the
Sommerfeld effect are very small ( $\sim \mathcal{O}(1)$) because
of the saturation behavior of the Sommerfeld effect. 
For the strongly resonant case the boost factor is typically smaller 
than $\sim \mathcal{O}(10)$. However, it is possible in some very 
extreme cases that DM distribution is adopted to give the maximal 
annihilation the boost factor can reach up to $\sim 1000$.
The variances of the boost factors due to different realizations of
substructures distribution are also discussed in the work.

\end{abstract}


\maketitle

\section{Introduction}

The recent observations of the positron excess by PAMELA
\cite{Adriani:2008zr} and the electron excess by ATIC
\cite{Chang:2008zz} and PPB-BETS \cite{Torii:2008xu} have
stimulated extensive discussions. The proposed explanations
include both the classical astrophysical processes (e.g.,
\cite{astro}) and possible new physics like dark matter (DM)
annihilation or decay (e.g. \cite{darkmatter}). For the annihilating 
DM scenario a ``boost factor'' of order $1000$ of the annihilation
rate, compared to that to produce the correct relic DM density
thermally at the early Universe, is needed to produce enough
electrons and positrons to fit the data. Although the newly published 
Fermi result on the electron spectrum \cite{fermi} does not
reproduce the ATIC sharp ``bump'' at $300 - 800$ GeV, however, to
explain the excess at Fermi by DM annihilation requires a similar
boost factor \cite{berg}.

There are several mechanisms suggested to generate the boost factor, 
including the DM substructures or DM mini-spikes \cite{Hooper:2008kv,
Bringmann:2009ip,Brun:2009aj}, nonthermal production of DM \cite{zhang}, 
the Sommerfeld effect (SE, \cite{sommerfeld,
Cirelli:2007xd,ArkaniHamed:2008qn,Pospelov:2008jd,Lattanzi:2008qa})
and the Breit-Wigner resonance enhancement \cite{breit}. The boost
factor due to DM substructures is a natural expectation as N-body
simulation shows that there are a large amount of substructures
exist in the Milky Way (MW) halo \cite{subhalo}. However, the
detailed calculation based on the N-body simulation shows that the
boost factor from DM substructures is generally less than $\sim10$
(\cite{Lavalle:1900wn}, hereafter we refer it as Paper I).
Although a single DM clump which is close enough to the Earth may
be possible to give large boost factor, it is found such a case
has very small probability to survive in a realistic DM
distribution model \cite{Lavalle:2006vb,Brun:2009aj}. Therefore,
additional boost effect like the SE or Breit-Wigner resonance
effect is still necessary in a DM picture to account for the
observational data.

Therefore it is necessary to extend the previous discussions of Paper 
I on the boost factor due to DM clumps to take into account the SE. 
Furthermore, such a study is also very important for the
indirect searches of DM. The SE or Breit-Wigner enhancement is
generally related with the velocity dispersion of DM particles.
Comparing with the MW halo, the DM in smaller structures is colder
and can give larger annihilation signals \cite{Pieri:2009zi,Bovy:2009zs}. 
In addition, if the DM substructures can contribute a proper fraction 
to the locally observed electrons/positrons, the boost factor of the
Galactic center (GC) can be suppressed due to the tidal destroy of
substructures in the inner Galaxy, and can avoid the strong constraints 
from $\gamma$-ray and radio emission \cite{Bertone:2008xr,Zhang:2008tb,
Bergstrom:2008ag,Cirelli:2009vg}.

In Refs. \cite{Pieri:2009zi,Bovy:2009zs} the SE in dwarf galaxies
and DM subhaloes has been investigated. However, in their works
only the effect on the source luminosity is considered. The
propagation of the charged particles, especially the most relevant
electrons/positrons, is not included. In this work, we will study
the boost factor of DM substructures on the electron/positrons and
antiprotons after incorporating the SE. 

To be clear, in the following part of this paper we will refer to the 
enhancement effect of DM substructures with respect to the smooth 
component as the ``boost factor'' (see definition in Sec. V). For the 
SE induced enhancement of the smooth component compared with the 
thermal production cross section we will directly call it as 
``SE enhancement''.

The paper is organized as follows. We first describe the SE in the
next section. Then we discuss the configurations of the DM 
substructures in Sec. III. The propagation models and definition of
boost factors are simply addressed in Sec. IV. In Sec. V we discuss
the results of the boost factors from DM substructures including SE. 
Finally we give the summary and conclusion.

\section{The Sommerfeld effect}

In the calculation of the DM annihilation cross section at low kinetic
energy regime, some non-perturbative effects may arise due to new
``long-range'' interaction. It is called ``Sommerfeld effect'' which
requires to sum all ladder diagrams due to the exchange of some
light scalars or gauge bosons between two incoming DM particles.
This effect could lead to significant enhancement of the
annihilation cross section at small relative velocity. To a good
approximation, one can simply multiply a factor $S$ to the
tree-level annihilation cross section. In the calculation of such
factor, we use the simplified quantum mechanical method in
literatures by solving the radial Schr\"{o}dinger equation with a
Yukawa potential $V(r)=-\frac{\alpha}{r} e^{-m_\phi r}$
\cite{Cirelli:2007xd,ArkaniHamed:2008qn}
\begin{equation}
\frac{1}{m_\chi}\psi''(r)+\frac{\alpha}{r}e^{-m_\phi
r}\psi(r)=-m_\chi \beta^2 \psi(r)\label{radial},
\end{equation}
where $\psi(r)$ is the reduced two-body wave function, $m_\chi$ and
$m_\phi$ are the masses of DM and the new mediating boson respectively,
$\beta$ is the velocity of DM in the center-of-mass frame. In this work,
we only consider the simplest situation in which the new mediator is a
light scalar or Abelian gauge boson. In addition, we only take into
account the incoming DM particles with $S$ wave (calculation for arbitrary
\textit{l}th partial wave could be found in Refs. \cite{Iengo:2009ni,
Cassel:2009wt}). 


As described in Ref. \cite{ArkaniHamed:2008qn}, there are two
equivalent methods to achieve $S$: (1) solving the Schr\"{o}dinger
equation with outgoing boundary condition
$\psi'(r)/\psi(r)\rightarrow im_\chi \beta$ as $r\rightarrow
\infty$, then $S$ is given by $S=|\psi(\infty)|^2/|\psi(0)|^2$;
(2) using boundary condition $\psi(r)/r\rightarrow
\textit{constant}$ as $r\rightarrow 0$, $S$ is given by
$S=|\psi'(0)/k|^2$, where $k$ is the momentum defined as
$k=m_\chi\beta$. Generally Eq.(\ref{radial}) needs to be solved
numerically.

In this work, we follow the method of Ref. \cite{Iengo:2009ni} to get 
$S$ numerically. We substitute $x=kr$ and $\phi(x)=C\psi(r)/kr$ into
Eq. (\ref{radial}) ($C$ is a constant), then solve it with boundary
condition as $\phi(0)=1$. We can normalize the solution at
infinity as $F(x)\equiv x\phi(x)\rightarrow C\cdot
\sin(x+\delta)$, and we have $C^2=F^2(x)+F^2(x+\pi/2)$ when $x$ is
large enough. Using the formula $S=|\psi'(0)/k|^2$, the SE
enhancement factor is obtained as $1/C^2$.

To achieve a more realistic result, we also need to take into account 
the speed distribution of DM particles \cite{Robertson:2009bh,Bovy:2009zs}.
After simply choosing Maxwell-Boltzmann distribution as an
acceptable approximation, the averaged SE enhancement is given by
\begin{equation}
\bar{S}=\frac{x}{K_2(x)}\int_0^{\infty}{\rm d}p'p'^2
\exp\left(-x\sqrt{1+p'^2}\right)S(\beta), \label{ava1}
\end{equation}
where $x$ is defined as $x=m_\chi/T$, $K_2(x)$ denotes the second order
modified Bessel function of the second type, $p'=\beta/\sqrt{1-\beta^2}$
is the normalized DM momentum. When the DM particles are non-relativistic
in the MW, Eq.(\ref{ava1}) can be written as
\begin{equation}
\bar{S}=\sqrt{\frac{2}{\pi}}\frac{1}{\sigma^3}\int_0^{v_{esc}}{\rm d}v
v^2\exp\left(-\frac{v^2}{2\sigma^2}\right)S(v),\label{ava2}
\end{equation}
where $v$ and $\sigma$ are the velocity and velocity dispersion of
DM respectively, $v_{esc}$ is the escape velocity in the MW. It is
obvious that if the behavior of $S$ are $S\sim 1/v$ and $S\sim 1/v^2$,
the average SE enhancement will have similar form
$\bar{S}\sim 1/\sigma$ and $\bar{S}\sim 1/\sigma^2$ respectively
\cite{Bovy:2009zs}.

\begin{figure}[htb]
\begin{center}
\includegraphics[width=0.45\columnwidth]{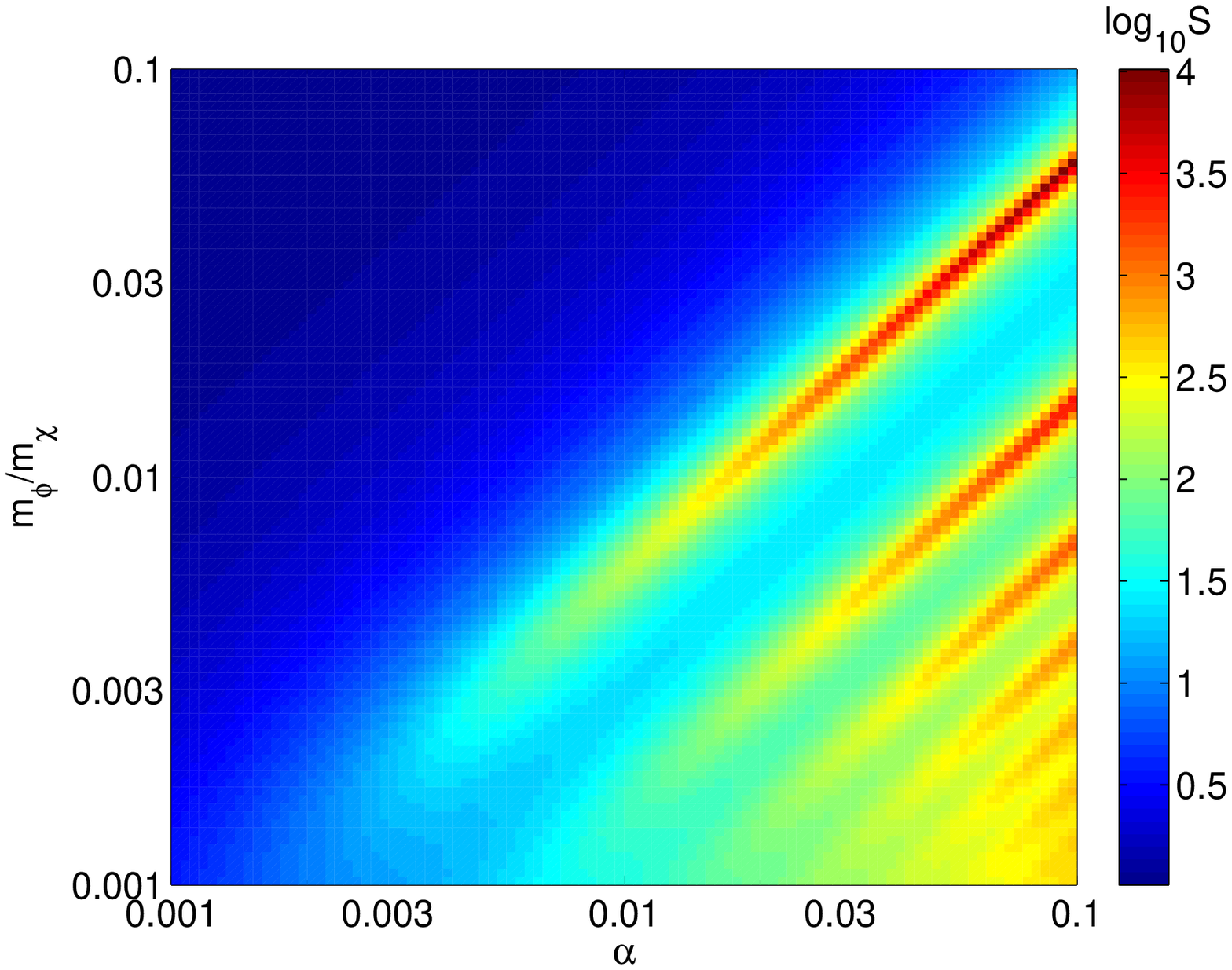}
\includegraphics[width=0.45\columnwidth]{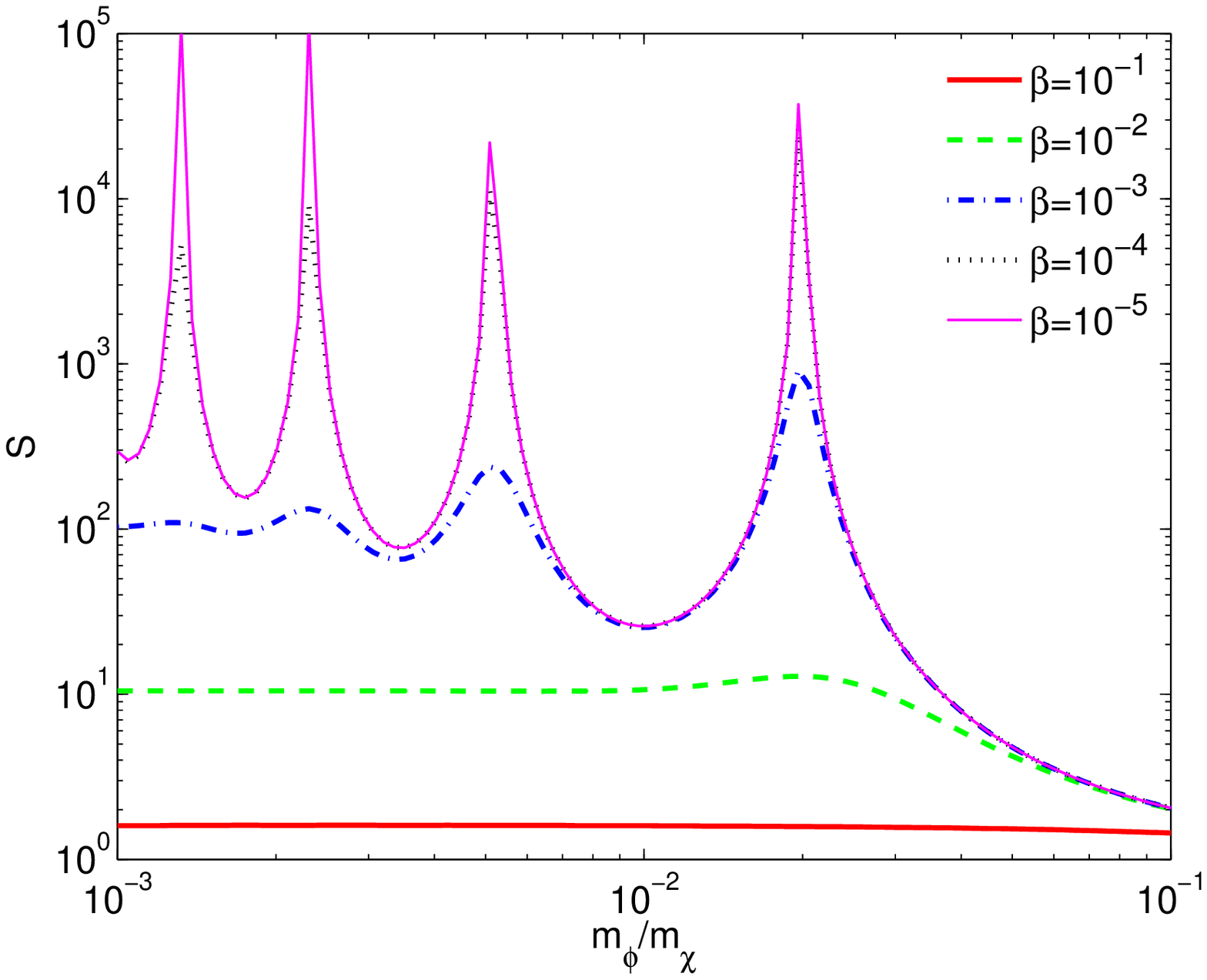}
\includegraphics[width=0.45\columnwidth]{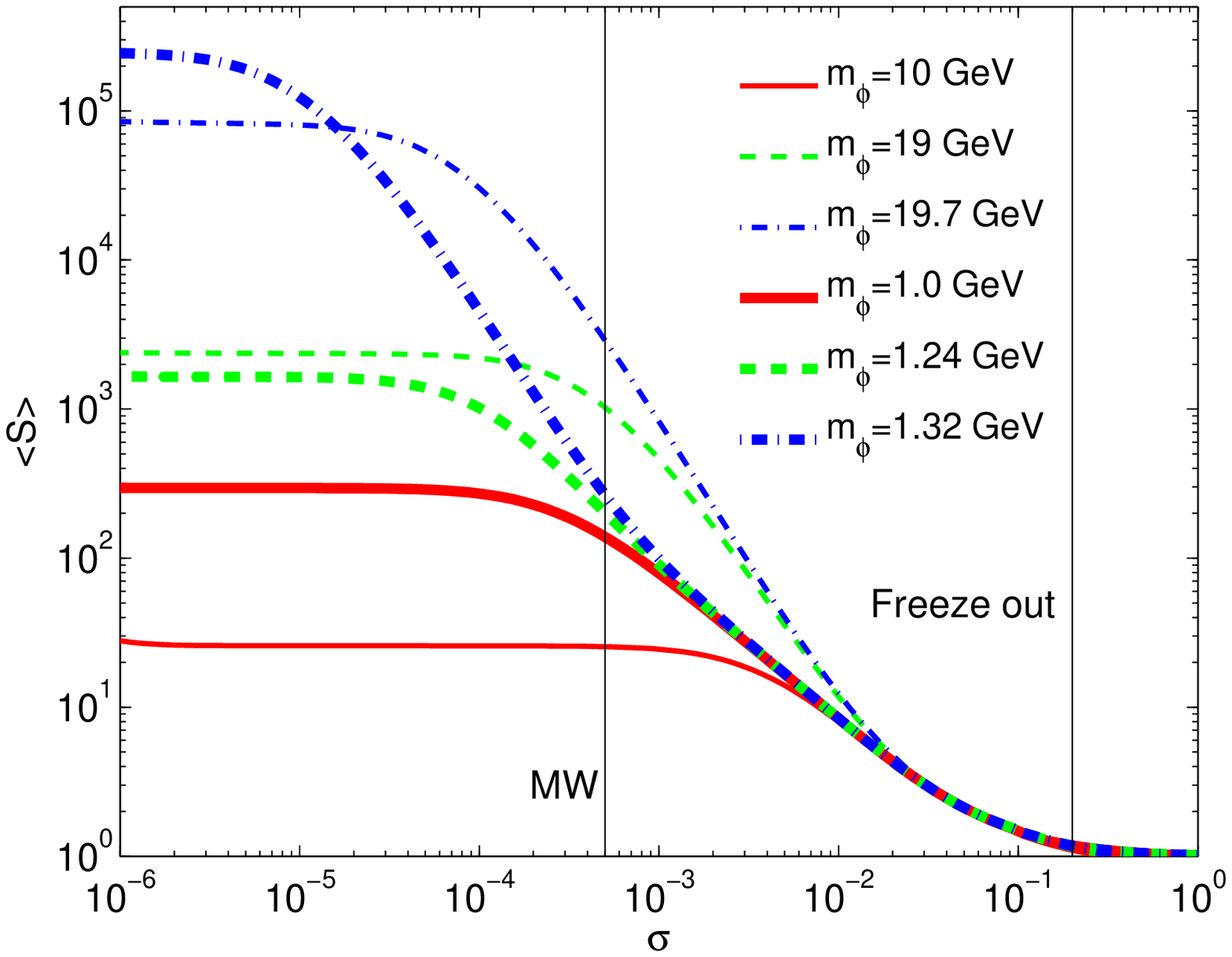}
\includegraphics[width=0.45\columnwidth]{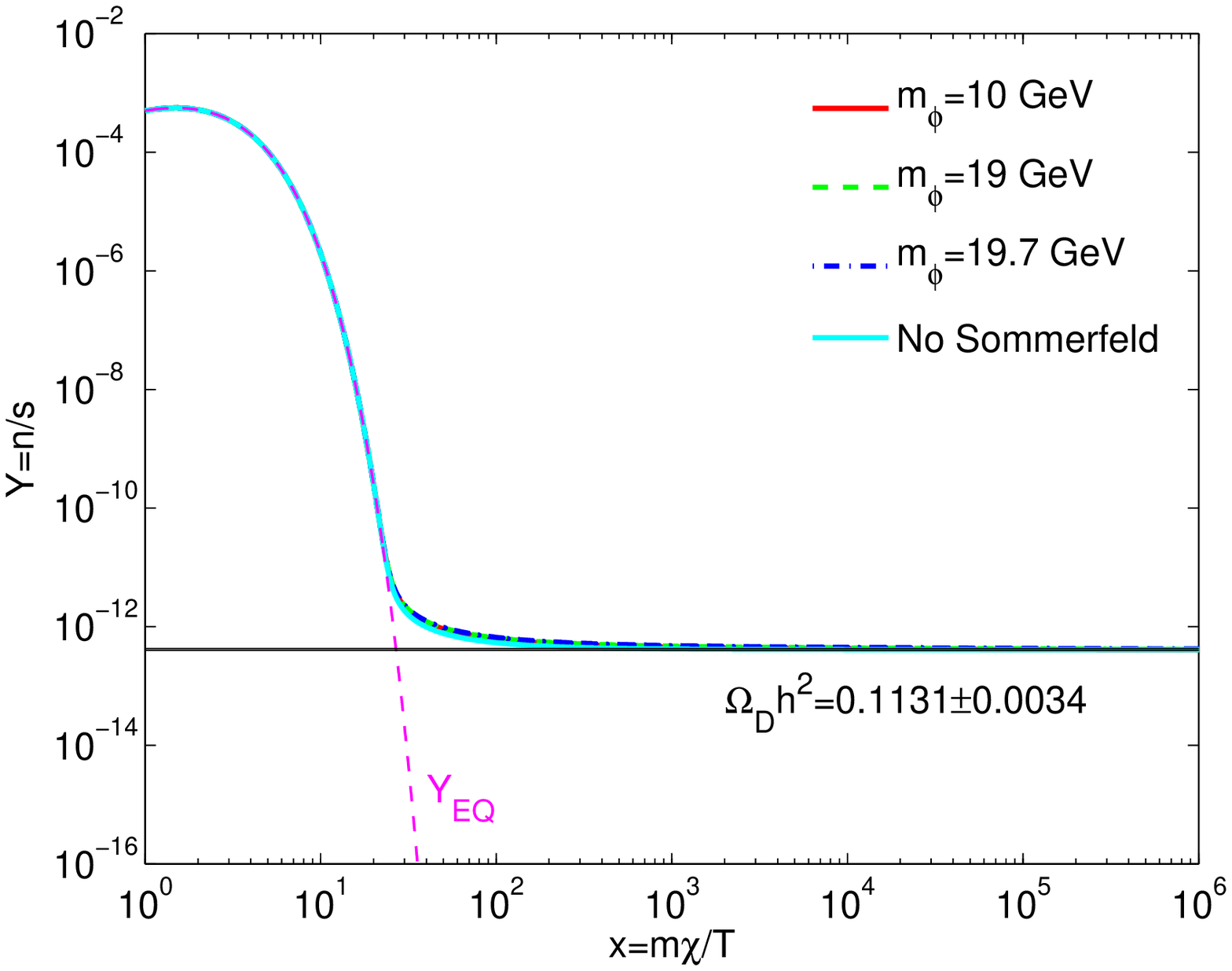}
\caption{Top-left: the SE enhancement factor $S$ as a function of
coupling constant $\alpha$ and $m_{\phi}/m_{\chi}$, for a single velocity
$\beta=220$ km s$^{-1}$; top-right: $S$ as a function of $m_{\phi}/m_{\chi}$
for several velocity $\beta$, in which $\alpha$ is fixed to be $1/30$;
bottom-left: the average SE enhancement factor vs. velocity
dispersion (in unit of light speed) of DM particles, for $\alpha=1/30$ and
$m_{\chi}=1$ TeV; bottom-right: the evolution of DM abundance with the
cosmic time $x\equiv m_{\chi}/T$ taking into account the SE corresponding
to the models in the bottom-left panel.
}
\label{fig:sommer}
\end{center}
\end{figure}

Some illustrations of the SE enhancements are shown in
Fig. \ref{fig:sommer}. In the upper two panels the variations of the
pre-averaged enhancement factor $S$ with model parameters are shown.
The results are similar with that given in Refs. \cite{Pieri:2009zi,
Bovy:2009zs}. From the top-right panel we can see that there are resonant
structures, especially for small velocity $\beta$. If the parameters are
tuned to approach the resonant region, very large boost factors can be
achieved. In the bottom-left panel, we show the average SE
enhancement factor $\bar{S}$ as a function of the velocity dispersion,
for several values of $m_{\phi}$: close to or away from the resonance
peak. In this calculation, we fix the mass of DM particle as $m_{\chi}=1$
TeV, and the coupling constant in the Yukawa potential as $\alpha=1/30$.
Two series of parameter $m_{\phi}$, around the two peaks $\sim 20$ GeV
and $\sim 1.3$ GeV, are shown in this panel. We can clearly see the 
``temperature'' dependence of the SE from this figure. It is shown 
that at the early stage of the evolution of the universe, when the 
DM is hot, the SE enhancement is negligible. When DM particles cool 
down, $\bar{S}$ become larger, and finally reach a platform when the 
DM particles are cold enough. We also label the velocity dispersion 
of today's MW halo, $\sigma\approx 5\times 10^{-4}$ ($150$ km s$^{-1}$). 
It can be inferred that if the difference of the SE between the 
smooth MW halo and the saturation value is larger, the boost effect 
from substructures should be more remarkable. We will go details on 
this point in Sec. V. In the bottom-right panel of Fig. \ref{fig:sommer} 
we show the evolution of DM abundance with the cosmic time $x\equiv 
m_{\chi}/T$ for the
models with $m_{\phi}=10,\,19$ and $19.7$ GeV shown in the bottom-left
panel. We find that there is almost no difference compared with the case
without SE. The annihilation cross section without SE, i.e.
$\langle\sigma v\rangle_0\approx 1.2-1.5\times 10^{-26}$ cm$^{3}$ s$^{-1}$
is found to give the right relic density of DM \cite{Komatsu:2008hk}
for different models.

In the following, we will employ the models shown in the
bottom-left panel of Fig. \ref{fig:sommer} to discuss the effects
on DM substructures from SE. These models cover the non-resonant
($m_\phi=10$ or $1$ GeV), moderately resonant ($m_\phi=19$ or
$1.24$ GeV) and strongly resonant ($m_\phi=19.7$ or $1.32$ GeV)
Sommerfeld enhanced cases without loss of generality. For other
parameters the conclusions can be easily translated. Note that in
this work we will focus on the boost factors from DM
substructures. We will not go in details of the particle physical
model of DM or the comparison of the expected fluxes of
$e^{\pm}$ and $\bar{p}$ with the data. However, some rough
implications from the observational data are adopted, like the
mass of DM $m_{\chi}\approx 1$ TeV according to ATIC
\cite{Chang:2008zz} or Fermi \cite{fermi} results. Actually as studied
in many related papers \cite{darkmatter,Zhang:2008tb}, a DM model
with $m_{\chi}\approx 1$ TeV and a total enhancement factor of about
several hundred can reproduce the observational data. That is to say
most of the choosen models as shown in the bottom-left panel of Fig. 
\ref{fig:sommer} (except for $m_{\phi}=10$ GeV) are potential ones 
which can explain the data, depending on the clumpiness boost factors 
that will be discussed below.

\section{DM distribution and substructures}

\subsection{Density profile}

The DM density profile based on N-body simulations can be
generally parameterized as a scale-invariant form
\begin{equation}
\rho=\frac{\rho_{\rm s}}{(r/r_{\rm s})^\gamma[1+(r/r_{\rm s})^\alpha]
^{(\beta-\gamma)/\alpha}},
\label{profile}
\end{equation}
where $\rho_{\rm s}$ and $r_{\rm s}$ are the scale density and radius
respectively, and $(\alpha,\,\beta,\,\gamma)$ are the shape parameters
which can be fitted from simulations. The simulations usually favor
a central cusp of the density profile, like the NFW profile with
$(\alpha,\,\beta,\,\gamma)=(1,3,1)$ \cite{Navarro:1996gj} and the Moore
profile with $(\alpha,\,\beta,\,\gamma)=(1.5,3,1.5)$ \cite{Moore:1999gc},
though the exact slope near the center is still under debate. In this
work we will employ both the NFW and Moore profiles for discussion.
Note that the central density for these profiles is divergent. To avoid
the singularity, we introduce a maximum central density $\rho_{\rm max}$
due to the fact that there should be a balance between the annihilating
rate and the in-fall rate of DM \cite{Berezinsky:1992mx}. For typical
parameter settings we have $\rho_{\rm max}= 10^{18}\sim10^{19}$ M$_{\odot}$
kpc$^{-3}$ (Paper I). Throughout this paper we fix
$\rho_{\rm max}$ to be $10^{18}$ M$_{\odot}$ kpc$^{-3}$ except special
claims.

The scale parameters $\rho_{\rm s}$ and $r_{\rm s}$ are determined using
the virial mass $M_{\rm vir}$, and the concentration parameter $c_{\rm vir}$.
The virial radius of a DM halo is defined as \begin{equation}
r_{\rm vir}=\left(\frac{M_{\rm vir}}{(4\pi/3)\Delta\rho_c}\right)^{1/3},
\label{rv}
\end{equation}
where $\Delta\approx 18\pi^2+82x-39x^2$ with $x=\Omega_M(z)-1=
-\frac{\Omega_{\Lambda}}{\Omega_M(1+z)^3+\Omega_{\Lambda}}$ (valid for
$\Lambda$CDM universe, \cite{Bryan:1997dn}) is the overdensity, and
$\rho_{\rm c}\approx 138$M$_{\odot}$ kpc$^{-3}$ is the critical density
of the universe. The concentration parameter $c_{\rm vir}$ is defined as
\begin{equation}
c_{\rm vir}=\frac{r_{\rm vir}}{r_{-2}},
\label{cv}
\end{equation}
where $r_{-2}$ refers to the radius at which $\frac{{\rm d} \left(
r^2\rho \right)}{dr} |_{r=r_{-2}}=0$. The concentration parameter
$c_{\rm vir}$ relates $r_{\rm vir}$ and the density profile parameter
as (Paper I)
\begin{equation}
r_s^{\rm NFW}=\frac{r_{\rm vir}(M_{\rm vir})}{c_{\rm vir}(M_{\rm
vir})},\ \ r_s^{\rm Moore}=\frac{r_{\rm vir}(M_{\rm
vir})}{0.63\,c_{\rm vir} (M_{\rm vir})}\ . \label{rs}
\end{equation}
Therefore if the $c_{\rm vir}-M_{\rm vir}$ relation is specified, $r_{\rm
s}$ is determined using Eq.(\ref{rs}). Finally we normalize the total
mass $\int\rho(r){\rm d}V$ to $M_{\rm vir}$ to get the scale density
$\rho_{\rm s}$.

Generally the $c_{\rm vir}-M_{\rm vir}$ relation can be derived through
fitting the observational mass profiles of gravitational systems like
galaxy clusters \cite{Buote:2006kx,Comerford:2007xb}. However, the
observational sample is limited in a narrow mass range and prevents us
from investigating the case down to very low mass haloes. Thus similar as
in Paper I we will use the toy model predictions on the
$c_{\rm vir}-M_{\rm vir}$ relation based on simulations. Two concentration
models B01 \cite{Bullock:1999he} and ENS01 \cite{Eke:2000av} are adopted
for discussion, with fitted polynominal form at $z=0$ as
\begin{equation}
\ln(c_{\rm vir})=\sum_{i=0}^4C_i\times\left[\ln\left(\frac{M_{\rm vir}}
{M_{\odot}}\right)\right]^i,
\label{lncv}
\end{equation}
with
\begin{equation}
C_i^{B01}=\{4.34,-0.0384,-3.91\times10^{-4},-2.2\times10^{-6},
-5.5\times10^{-7}\},
\label{cvb01}
\end{equation}
and
\begin{equation}
C_i^{ENS01}=\{3.14,-0.018,-4.06\times10^{-4},0,0\}.
\label{cvens01}
\end{equation}

\subsection{Substructures}

The mass and spatial distributions of DM substructures can be parameterized
as \cite{Diemand:2004kx,Diemand:2005vz}
\begin{equation}
\frac{{\rm d}N_{\rm sub}(r,M_{\rm sub})}{{\rm d}V{\rm d}M_{\rm sub}}
=N_{\rm sub}\times \frac{{\rm d}P_{\rm M}(M_{\rm sub})}{{\rm d}M_{\rm sub}}
\times \frac{{\rm d}P_{\rm V}(r)}{{\rm d}V},
\end{equation}
where $N_{\rm sub}$ is the total number of subhaloes, ${\rm d}P_{\rm M}/
{\rm d}M_{\rm sub}$ and ${\rm d}P_{\rm V}/{\rm d}V$ are the normalized
mass and spatial distribution probabilities. For the mass function, the
N-body simulations show a power law distribution
\begin{equation}
\frac{{\rm d}P_{\rm M}(M_{\rm sub})}{{\rm d}M_{\rm sub}}\propto
\left(\frac{M_{\rm sub}}{M_{\odot}}\right)^{-\alpha_{\rm m}}.
\label{dpdm}
\end{equation}
This relation is assumed to hold in a wide mass range, from the most 
massive subhalo in the MW, $M_{\rm max}\sim 10^{10}$ M$_{\odot}$, down 
to the scale of Earth mass $M_{\rm min}\sim 10^{-6}$ M$_{\odot}$
\cite{Diemand:2005vz}. The power law index $\alpha_{\rm m}$ is
about $2$, however, with a scattering from $\sim 1.7$ to $\sim
2.1$ in various works \cite{subhalo,Gao:2004au,Shaw:2005dy}. The
two recent highest resolution simulations, Aquarius and Via Lactea
find the values of $\alpha_{\rm m}$ to be $1.9$ and $2.0$
respectively \cite{Diemand:2006ik,Springel:2008cc}. For
$\alpha_{\rm m}\gtrsim 2.0$ the mass fraction of subhaloes will be
sensitively dependent on the lower cut of the mass of subhalo,
which is still very uncertain
\cite{Berezinsky:2003vn,Green:2003un,Profumo:2006bv}. In this work
we adopt the fiducial values of $\alpha_{\rm m}=1.9$ and the
minimum mass $M_{\rm min}=10^{-6}$ M$_{\odot}$ as benchmark model,
while the results of other parameters are also given for
comparison.

The spatial distribution of subhaloes is usually found to be anti-biased
with respect to the DM density distribution, and can be fitted with an
cored isothermal function (e.g.,\cite{Diemand:2004kx})
\begin{equation}
\frac{{\rm d}P_{\rm V}(r)}{4\pi r^2{\rm d}r}\propto\left[1+\left(
\frac{r}{r_{\rm H}}\right)^2\right]^{-1},
\label{dpdv}
\end{equation}
where $r_{\rm H}\approx 0.14 r_{\rm vir}^{\rm MW}$ is the core radius.

The normalization of the total number of subhaloes is determined by
setting the number with mass heavier than $10^8$ M$_{\odot}$ is $100$
(Paper I). For such a normalization and the mass function
slope $\alpha_{\rm m}=1.9$, we find the total number of subhaloes
with minimum mass $M_{\rm min}=10^{-6}$ M$_{\odot}$ is about
$4\times 10^{14}$, which is consistent with the one obtained in
Ref. \cite{Diemand:2005vz}.

Finally, we simply give the overall property of the MW halo. We will adopt
a NFW profile with total mass $M_{\rm MW} \approx 10^{12}$ M$_{\odot}$
\cite{Xue:2008se}. The virial radius of the MW halo is about $270$ kpc,
and the concentration parameter calculated using B01 model is about
$13.6$. The local density is then calculated to be $\rho_0(r_{\odot}=8.5
\,{\rm kpc})\approx 0.25$ GeV cm$^{-3}$. Note that a fraction of mass
$f$ will be in substructures, so the actual density of the so-called
``smooth'' component is $(1-f)\rho(r)$. For the above benchmark
configuration of subhaloes, $f\approx 0.14$ is found. Since we mainly
focus on the boost factors from subhaloes, this smooth halo model is
fixed in the following discussion.

\section{Propagation model and boost factors}

In the Galaxy the transport of charged particles is affected by
several processes. The scattering off random magnetic fields will
lead to spatial and energy diffusions. The stellar wind may also
blow away the cosmic rays (CRs) from the Galactic plane. In
addition, interactions of CR particles with the interstellar
radiation field (ISRF) and/or the interstellar medium (ISM) can
result in continuous and catastrophic energy losses. Since the
detailed processes affect the propagation are species-dependent,
we will describe the treatments for antiprotons and positrons
below respectively. The basic common framework is as follows. For
the transport processes we take a spatial independent diffusion
coefficient $D(E)=\beta D_0{\cal R}^{\delta}$ (where ${\cal
R}=pc/Ze$ is the rigidity) and a constant wind $V_{\rm c}$
directed outwards along $z$. CRs are confined within a cylinder
halo $L$, i.e. the differential density, $dN/dE\equiv n$, is bound
by $n(z=\pm L, R_{\rm max})=0$ with $R_{\rm max}$ of the scale of
the visible Galaxy\footnote{Note that the actual effective volume
of CRs depends on the propagation parameters, see e.g.,
\cite{Maurin:2002uc}.}. The free parameters of the model are the
halo size $L$ of the Galaxy, the normalization of the diffusion
coefficient $K_0$ and its slope $\delta$, and the constant
galactic wind $V_{\rm c}$.

The propagation equation of CRs can be generally written as
\begin{equation}
-D\Delta N+V_{\rm c}\frac{\partial N}{\partial z}+2h\Gamma_{\rm tot}
\delta(z)N
+ \frac{\partial}{\partial E}\left(\frac{{\rm d}E}{{\rm d}t}N\right)
= q({\bf x},E),
\label{prop}
\end{equation}
where $\Gamma_{\rm tot}=\sum_{i=H,He}n_i\,\sigma_i\,v$ is the
destruction rate of CRs through interaction with ISM in the thin
gas disk with half height $h\approx 0.1$ kpc, ${\rm d}E/{\rm d}t$
is the energy loss rate, and $q({\bf x},E)$ is the source function.

Given the propagated fluxes of CRs, we then define the boost factor 
as the ratio of the sum of the smooth and substructure contributions 
to the smooth one without substructures. 
The detailed formula of the solutions of the propagation equations 
and the boost factor are presented in the Appendix. For more 
details please refer to Paper I and references therein.

\section{Results}

For the convenience of comparison, we specify the reference model
configuration based on the descriptions in previous sections:
$M_{\rm min}=10^{-6}$ M$_{\odot}$, $\alpha_{\rm m}=1.9$; the inner
profile of subhalo is NFW, and the concentration model is B01; the
propagation parameters are the median settings (see Appendix A). 
We also include the discussion about an extreme configuration with 
$\alpha_{\rm m}=2.0$, Moore inner profile and B01 concentration model. 
These two parameter settings are compiled in Table \ref{table:sub}.

\begin{table}[!htb]
\begin{center}
{\begin{tabular}{cccccc}
\hline
\hline
 & $\alpha_{\rm m}$ & profile & $M_{\rm min}$($M_\odot$) & concentration & propagation  \\
\hline
reference & 1.9 & NFW   & $10^{-6}$ & B01 & MED   \\
extreme   & 2.0 & Moore & $10^{-6}$ & B01 & MED   \\
\hline
\end{tabular}}
\caption{The reference and extreme model configurations of DM substructures.
\label{table:sub}}
\end{center}
\end{table}

\begin{figure}[htb]
\begin{center}
\includegraphics[width=0.45\columnwidth]{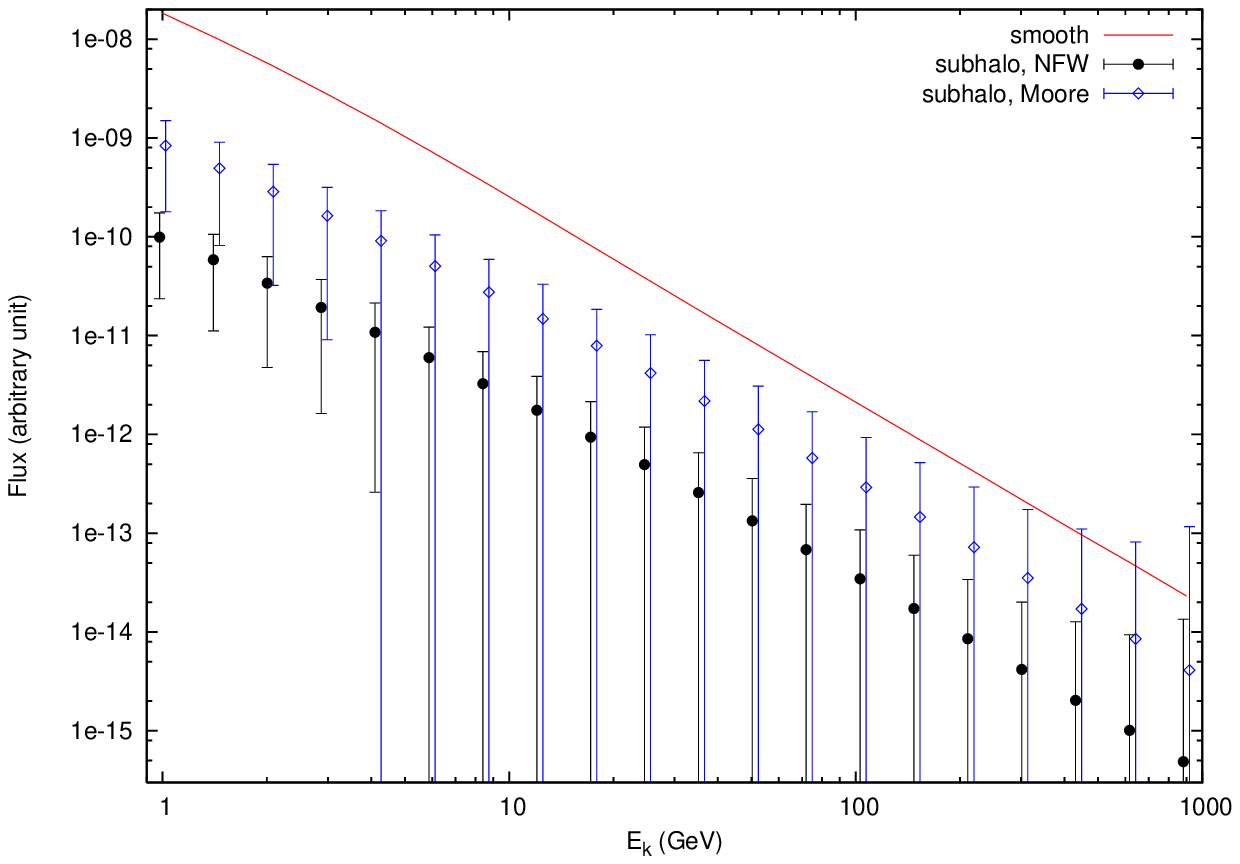}
\includegraphics[width=0.45\columnwidth]{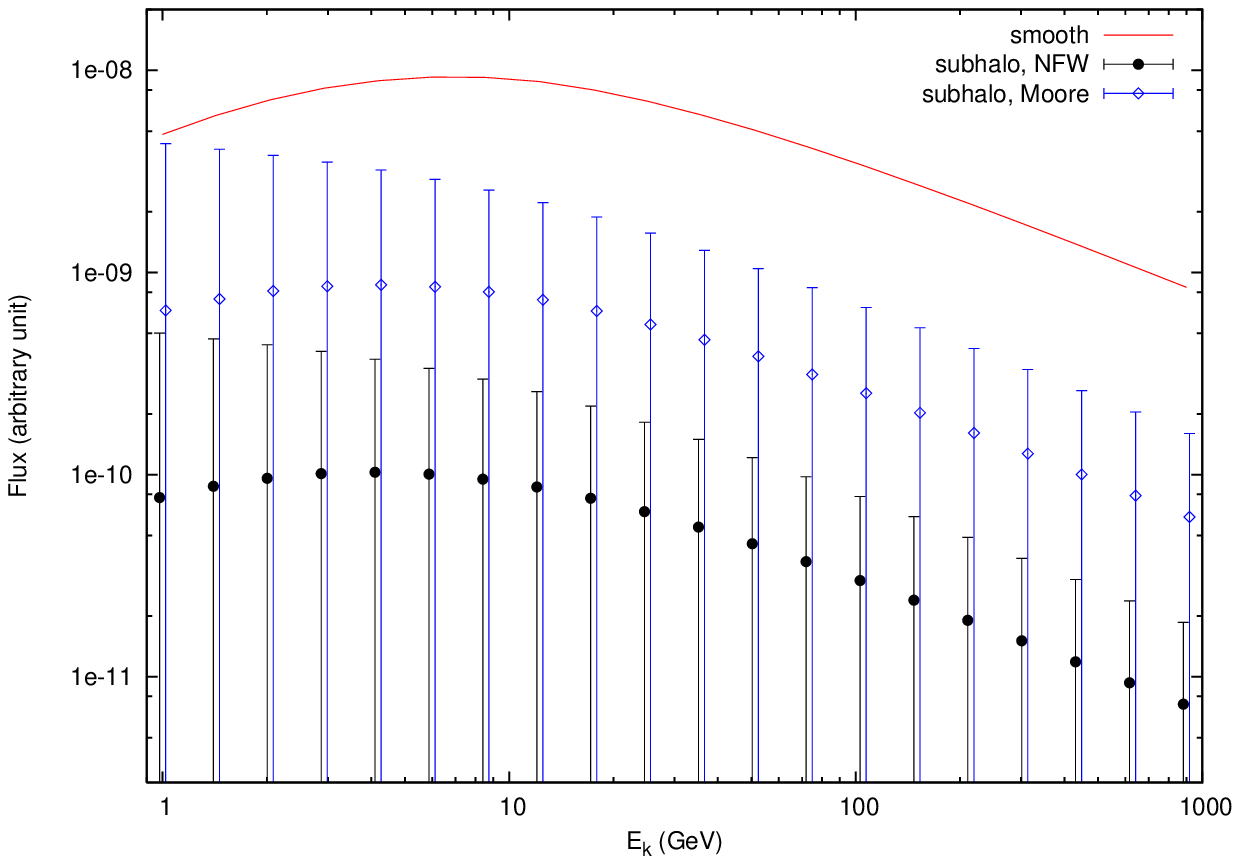}
\caption{The (relative) fluxes of positrons (left) and antiprotons 
(right) from DM annihilation for the smooth component (red solid line), 
the subhalo component with reference configuration (black full circle), 
and the subhalo component with Moore inner profile (blue empty diamond).} 
\label{fig:flux}
\end{center}
\end{figure}

To get a rough idea about how large the enhancement due to SE is necessary 
to give a non-negligible boost factor, we give the relative fluxes of 
positrons (left) and antiprotons (right) in the absense of SE in Fig. 
\ref{fig:flux}. It is shown that the contribution to the charged 
particle fluxes of DM subhaloes for the reference configuration is
about two orders of magnitude lower than the smooth component. If the 
inner profile of DM subhalo is as cuspy as Moore profile, the resulting
contribution from DM subhaloes is still about one order of magnitude
lower than the smooth one. Thus in the absence of SE case, it is very
difficult to generate large enough boost factor only from DM clumpiness.
More details were discussed in Paper I.

In the following we will discuss the cases including the SE. Three kinds 
of SE enhanced cases, i.e., the non-resonant case ($m_{\phi}=10$ or $1$ 
GeV of Fig. \ref{fig:sommer}), the moderately resonant case ($m_{\phi}=19$ 
or $1.24$ GeV) and the strongly resonant case ($m_{\phi}=19.7$ or $1.32$ 
GeV) respectively, are discussed one by one.

\subsection{Non-resonant case}

\begin{figure}[htb]
\begin{center}
\includegraphics[width=0.45\columnwidth]{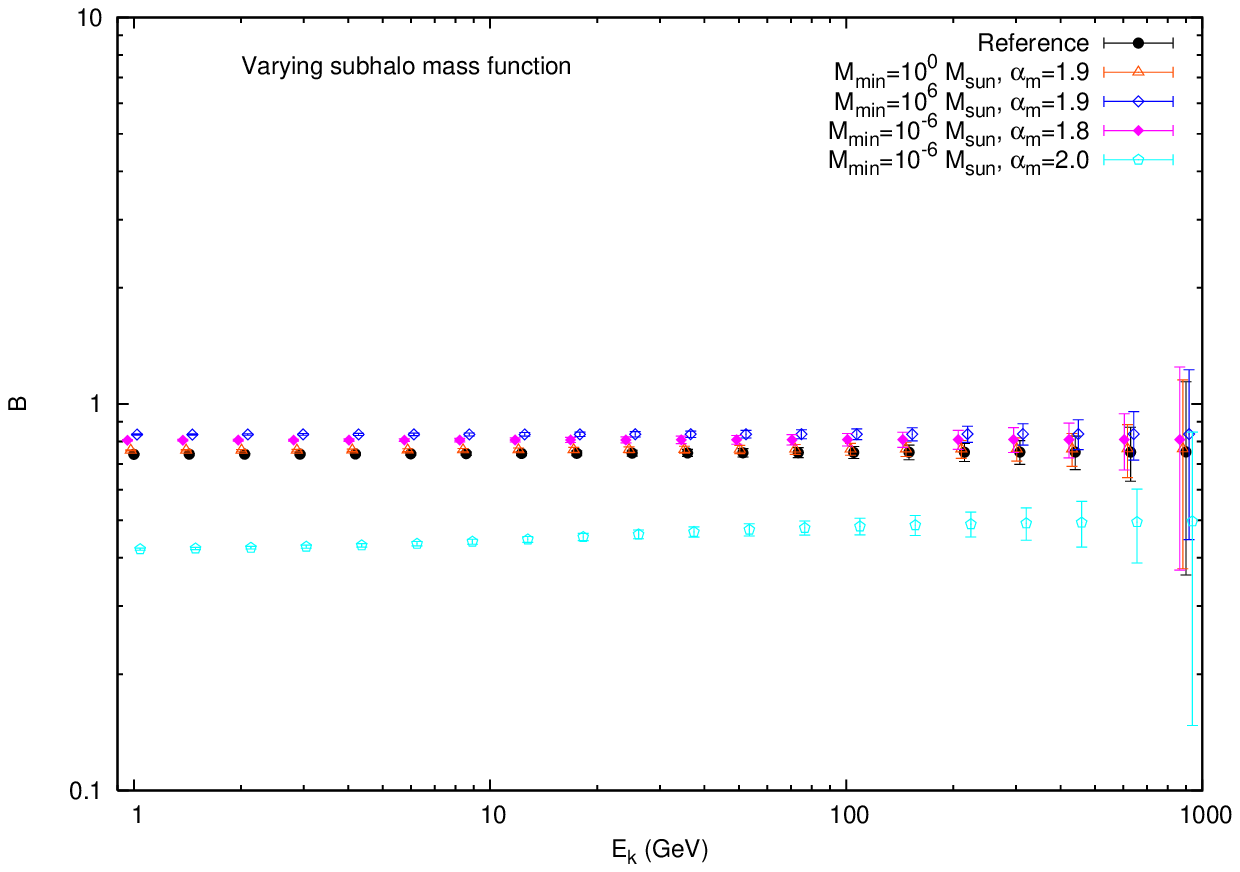}
\includegraphics[width=0.45\columnwidth]{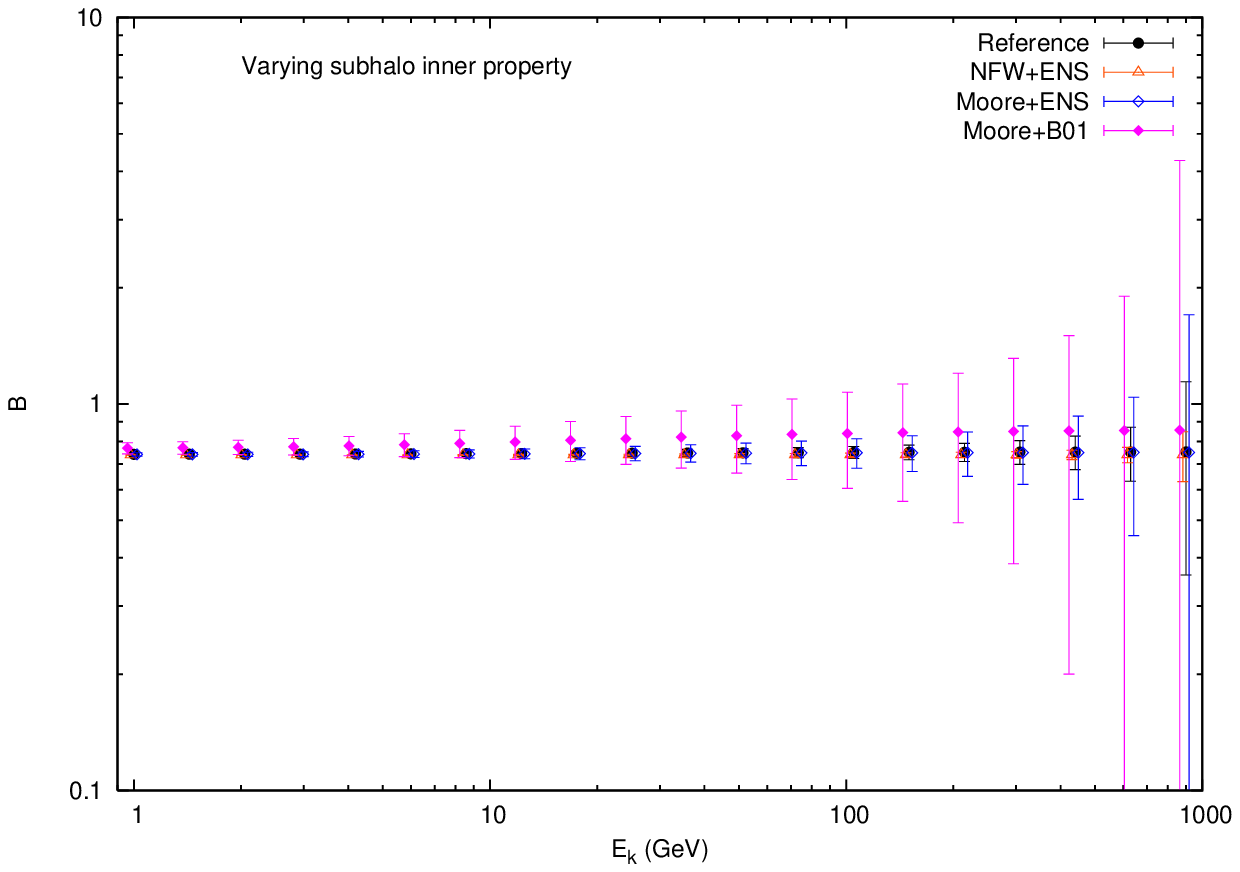}
\includegraphics[width=0.45\columnwidth]{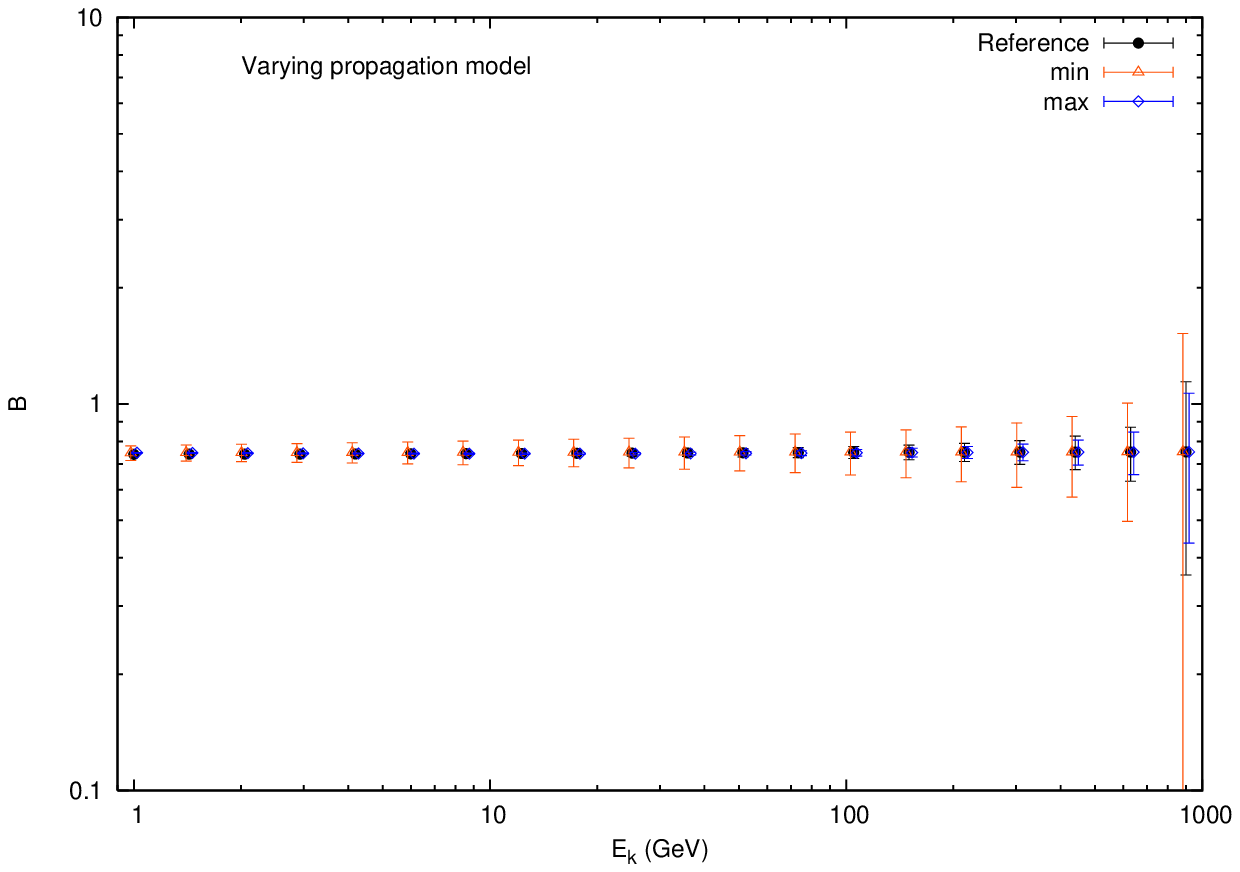}
\includegraphics[width=0.45\columnwidth]{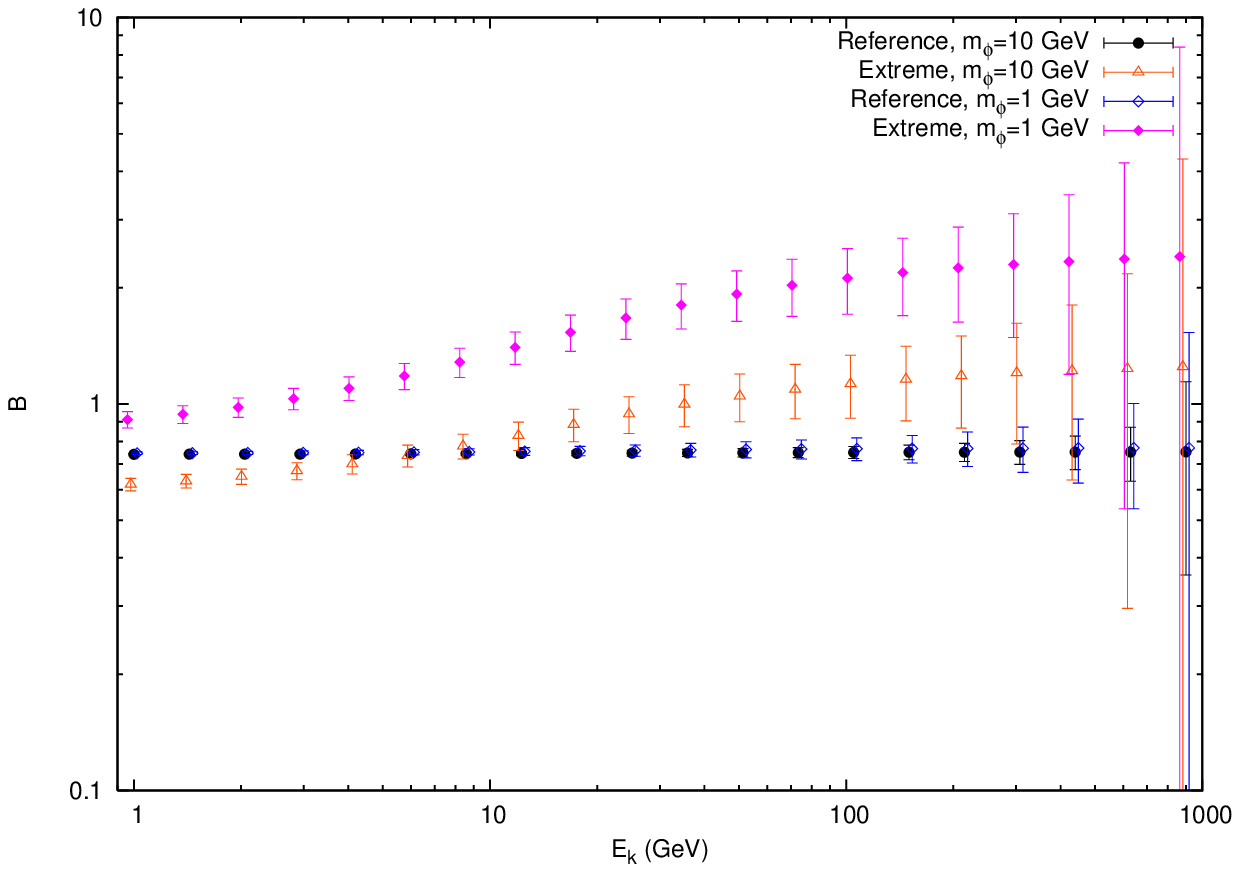}
\caption{The boost factors of positrons for the non-resonant SE
case, for different model configurations: varying the distribution
of subhalo population (top-left); varying the inner property of
subhalo (top-right); varying the propagation parameters
(bottom-left); and finally the extreme cases with $\alpha_{\rm
m}=2.0$, Moore inner profile with B01 concentration model
(bottom-right). The errorbars show the variances of the boost
factors. For clarity of the plot, we slightly shift the energy
axis among different models in one panel. } \label{fig:pos1}
\end{center}
\end{figure}

\begin{figure}[htb]
\begin{center}
\includegraphics[width=0.45\columnwidth]{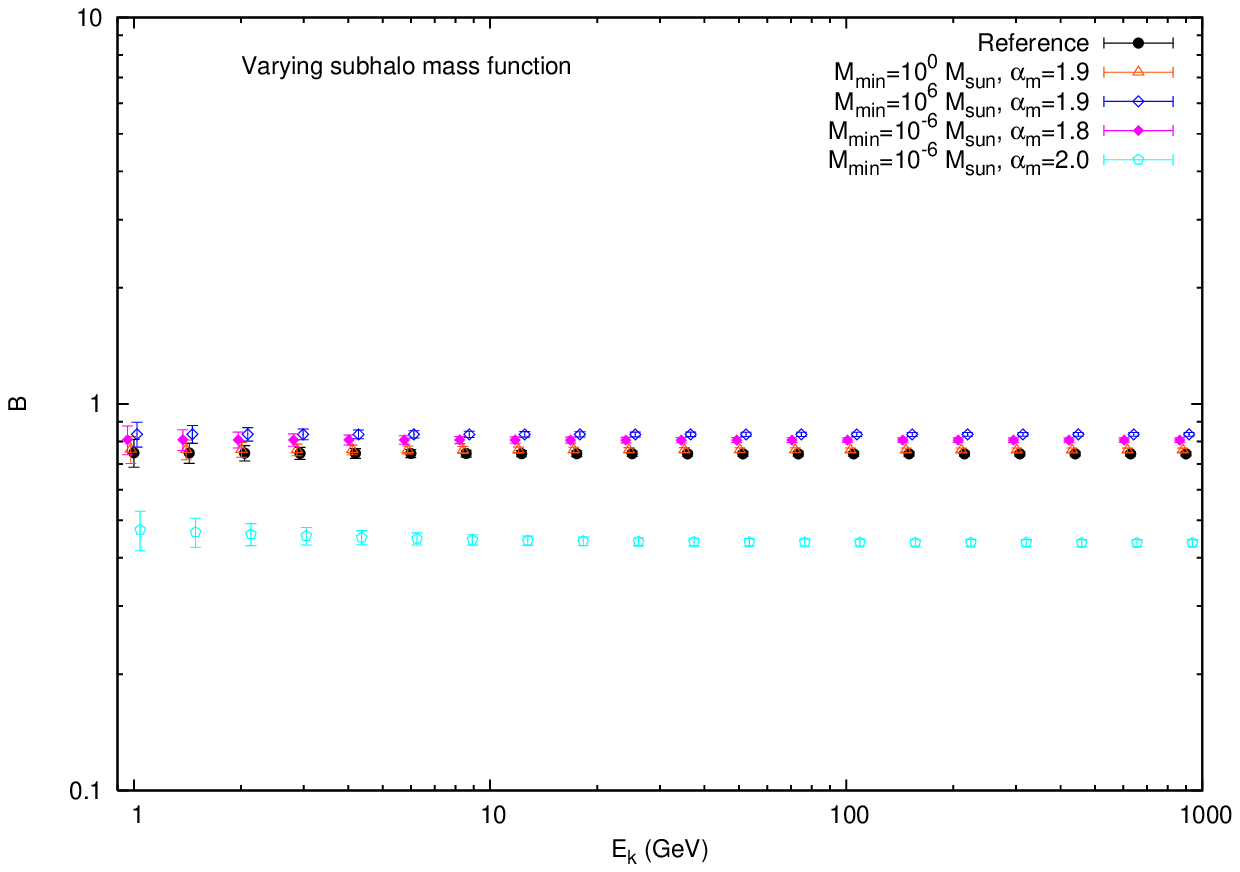}
\includegraphics[width=0.45\columnwidth]{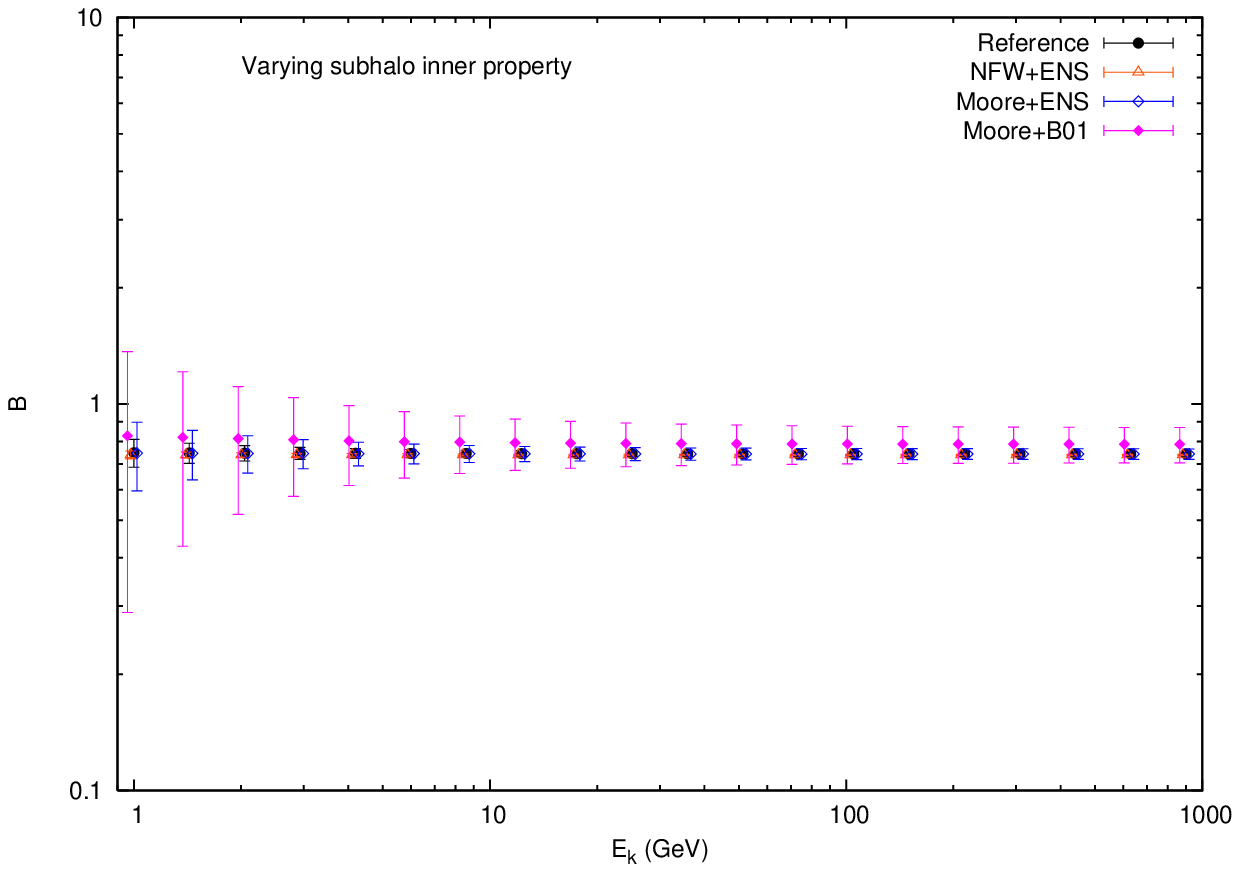}
\includegraphics[width=0.45\columnwidth]{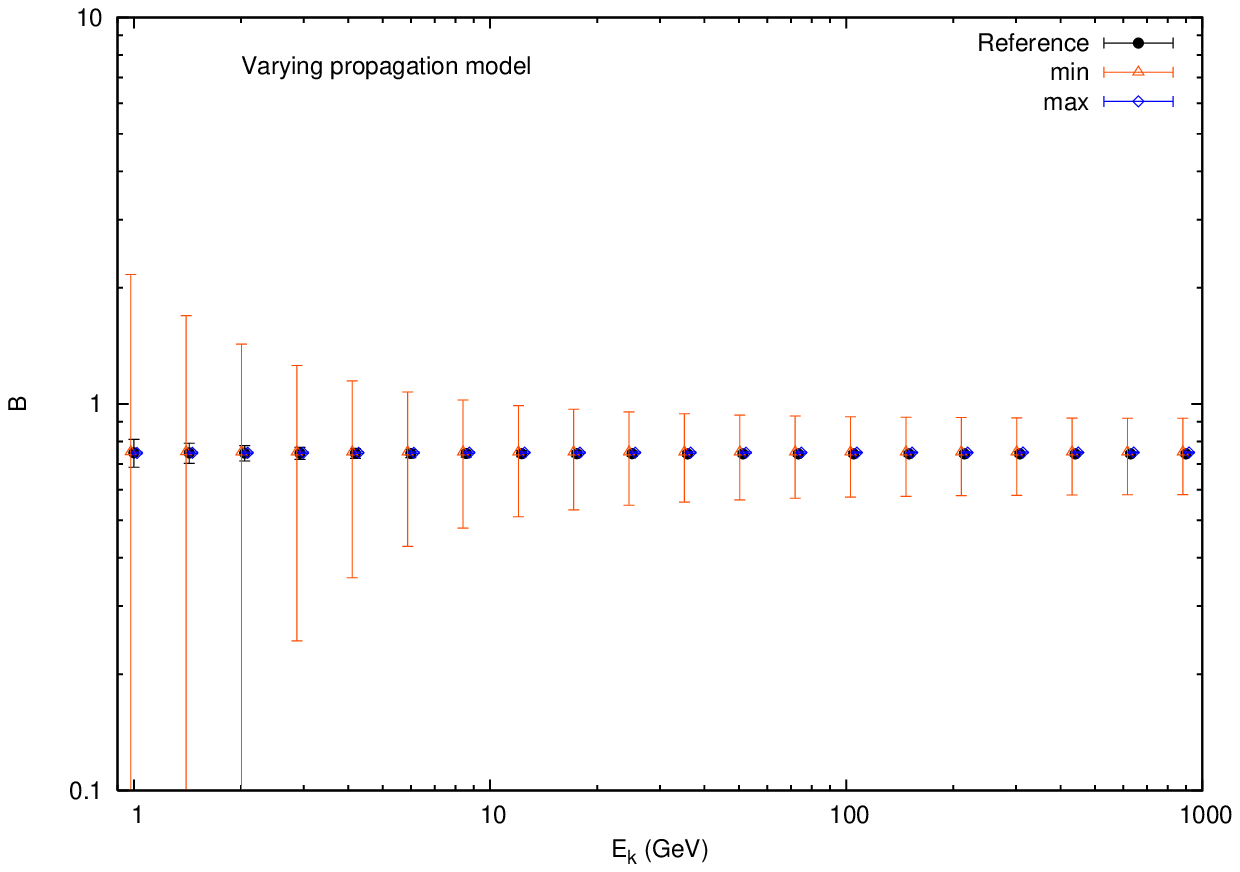}
\includegraphics[width=0.45\columnwidth]{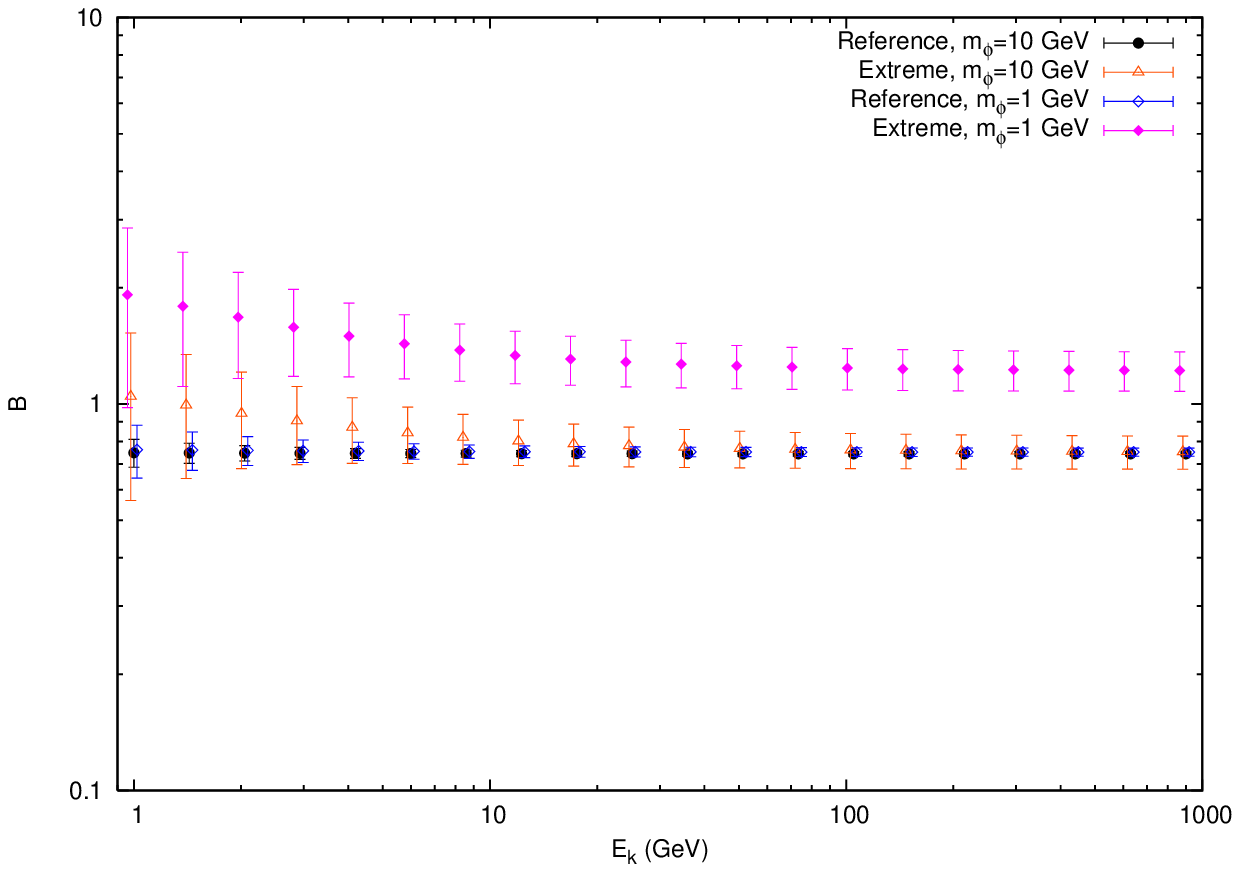}
\caption{The same as Fig. \ref{fig:pos1} but for antiprotons.
}
\label{fig:pba1}
\end{center}
\end{figure}

In Figs. \ref{fig:pos1} and \ref{fig:pba1} we show the boost
factors for $e^+$ and $\bar{p}$ respectively, for the non-resonant
SE model. For the first three panels discussing the effects of DM
distribution or propagation models we set $m_{\phi}=10$ GeV. Only
in the last panel we show the comparison between various values of
$m_{\phi}$. From Fig. \ref{fig:sommer} we can see that for
$m_{\phi}=10$ GeV, there should be no additional enhancement
effect on the substructures from SE because the smooth halo has
lay in the saturation region. In other words the SE enhancement
factor will be the same for both the smooth component and the
substructures. Therefore the results should be the same as
discussed in Paper I, where we find the boost effects from DM
substructures are negligible for DM distributions from N-body
simulations.

For the $m_{\phi}=1$ GeV case, we find a factor about $2$ times
larger for the saturation value than the smooth MW halo. Therefore
it will lead to $\sim 2$ times larger boost factor than that for
$m_{\phi}=10$ GeV since most of the subhaloes lie in the satuation
region\footnote{For the most massive subhalo $M_{\rm max}=10^{10}$
M$_{\odot}$, we have $\sigma\approx 2\times 10^{-4}$.}. This can
be seen from the extreme case\footnote{Note, however, since the
boost factor is dominated by the smooth component for the
reference configuration, we do not see remarkable differences
between $m_{\phi}=10$ and $1$ GeV for the reference model.} in the
last panels of Figs. \ref{fig:pos1} and \ref{fig:pba1}.

From the Figs. \ref{fig:pos1} and \ref{fig:pba1} we have more
details about the boost factor and its uncertainty.
\begin{itemize}

\item Energy dependence of the boost factor $B$.

As firstly pointed out in Ref. \cite{Lavalle:2006vb}, the boost
factor from DM substructures is energy dependent instead of a
constant due to the energy-dependent propagation effects of
charged CRs. This energy-dependent propagation can be further
translated into the differences of spatial distributions between
the smooth component and the substructures. This is because the
effective propagation lengths of $e^+$ and $\bar{p}$ vary with
energy \cite{Maurin:2002uc,Lavalle:1900wn}. Specificly, for $e^+$
the propagation length decreases with the increase of energy due
to faster energy loss of high energy positrons; while for
$\bar{p}$ the case is just contrary. Thus we may expect a lower
boost factor for low energy positrons since it reflects the ratio
of substructures to the smooth halo in a larger volume, which
includes more smooth contribution when closing to the GC.
Similarly, for antiprotons we will expect a higher boost factor
for low energy particles. These proterties can be seen clearly in
the bottom-right panels of Figs. \ref{fig:pos1} and
\ref{fig:pba1}.

\item Energy dependence of the variance $\sigma_{\rm B}$.

The variance of the boost factor $\sigma_{\rm B}$ is also related to the
effective propagation volume, and hence the energy, of CRs. The smaller
the propagation volume, the larger the statistical uncertainty. Therefore
the variance is larger for $e^+$ (smaller for $\bar{p}$) at high energies.
These behaviors are also shown in Figs. \ref{fig:pos1} and \ref{fig:pba1}.

\item Dependence on the subhalo mass function.

Since we use the number of subhalos above $10^8$ M$_{\odot}$ to 
normalize the number distribution function, the
lower mass cut $M_{\rm min}$ and the mass function slope
$\alpha_{\rm m}$ will affect the fraction of subhalo contribution
to the total signals. If $M_{\rm min}$ is smaller, or $\alpha_{\rm
m}$ is larger, the mass fraction of substructures $f$ will be
larger (see Paper I). However, when the boost from subhalos is not 
significant a large fraction of mass in subhalos may lead to suppress 
of DM annihilation, as shown in the figures. This is because the boost
factor approximates as $(1-f)^2$ according to Eq.(\ref{boost}).

It is not direct to derive the influence on the variance of the boost factor.
From Eq.(\ref{variance}) we know that $\sigma_{\rm B}\propto\sigma_{\rm sub}$
with $\sigma_{\rm sub}$ given in Eq.(\ref{rela_vari}). The variance of subhalo
flux will depend on the total number of subhaloes $N_{\rm sub}$, the average
and variance of the annihilation luminosity $\xi$ of a single clump. As
shown in Figs. \ref{fig:pos1} and \ref{fig:pba1} (and also the related
discussions in Paper I), $\sigma_{\rm B}$ does not depend sensitively on
the parameters $M_{\rm min}$ and $\alpha_{\rm m}$.

\item Dependence on the inner property of subhalo.

In the top-right panels of Figs. \ref{fig:pos1} and \ref{fig:pba1}
we show the boost factors together with the variances when
changing the subhalo inner profile and the concentration model. It
is known that B01 concentration model would give larger
annihilation luminosity $\xi$ than ENS model, and Moore profile
would also give larger flux than NFW profile. Therefore we get the
largest boost factor for Moore + B01 configuration, while the
smallest one for NFW + ENS configuration, although both are still
negligible ($\sim 1$). As for the variance, since the annihilation
luminosity $\xi$ for ENS concentration model or Moore profile
differs from the reference model (B01, NFW) by nearly a constant
factor\footnote{It is weakly dependent on clump mass $M_{\rm
sub}$, see Fig. 3 of Paper I.}, we expect
$\sigma_\xi/\langle\xi\rangle_M$ to be almost unchanged. Then from
Eq.(\ref{rela_vari}) we have $\sigma_{\rm sub}\propto
\langle\Phi_{\rm sub}\rangle$, which is the largest for Moore +
B01 model and smallest for NFW + ENS model.

\item Dependence on the propagation model.

The effects of various propagation parameters given in Table \ref{table:prop}
are shown in the bottom-left panels of Figs. \ref{fig:pos1} and
\ref{fig:pba1}. Since the propagation parameters affect both the smooth
component and substructures, the boost factor will not be sensitive to
the propagation parameters. However, the propagation parameters will strongly
affect the effective propagation length of particles ($\lambda_{\rm D}^{e^+}
\propto\sqrt{D_0}$, and $\lambda_{\rm D}^{\bar{p}}\propto D/V_{\rm c}$),
and then lead to different variances. For the minimum parameter setting
the effective propagation length is the smallest, so the variance is the
largest.

\item The $maximal$ case.

We now come to the extreme model configuration with all the
maximum settings of parameters discussed above, i.e., B01
concentration model, Moore inner profile and mass function slope
$\alpha_{\rm m}=2.0$\footnote{When the flux from subhaloes become
to dominate the smooth component, the model with $\alpha_{\rm
m}=2.0$ tends to give the larger boost factor than the one with
$\alpha_{\rm m}=1.8$.}, to show the $maximal$ boost factor. The
median propagation model is adopted due to the fact that the boost
factor does not depend on the propagation parameters sensitively.
We find that in this extreme case, the $maximal$ boost factor will
be less than $2\sim 3$.

\end{itemize}

\subsection{Moderately resonant case}

\begin{figure}[htb]
\begin{center}
\includegraphics[width=0.45\columnwidth]{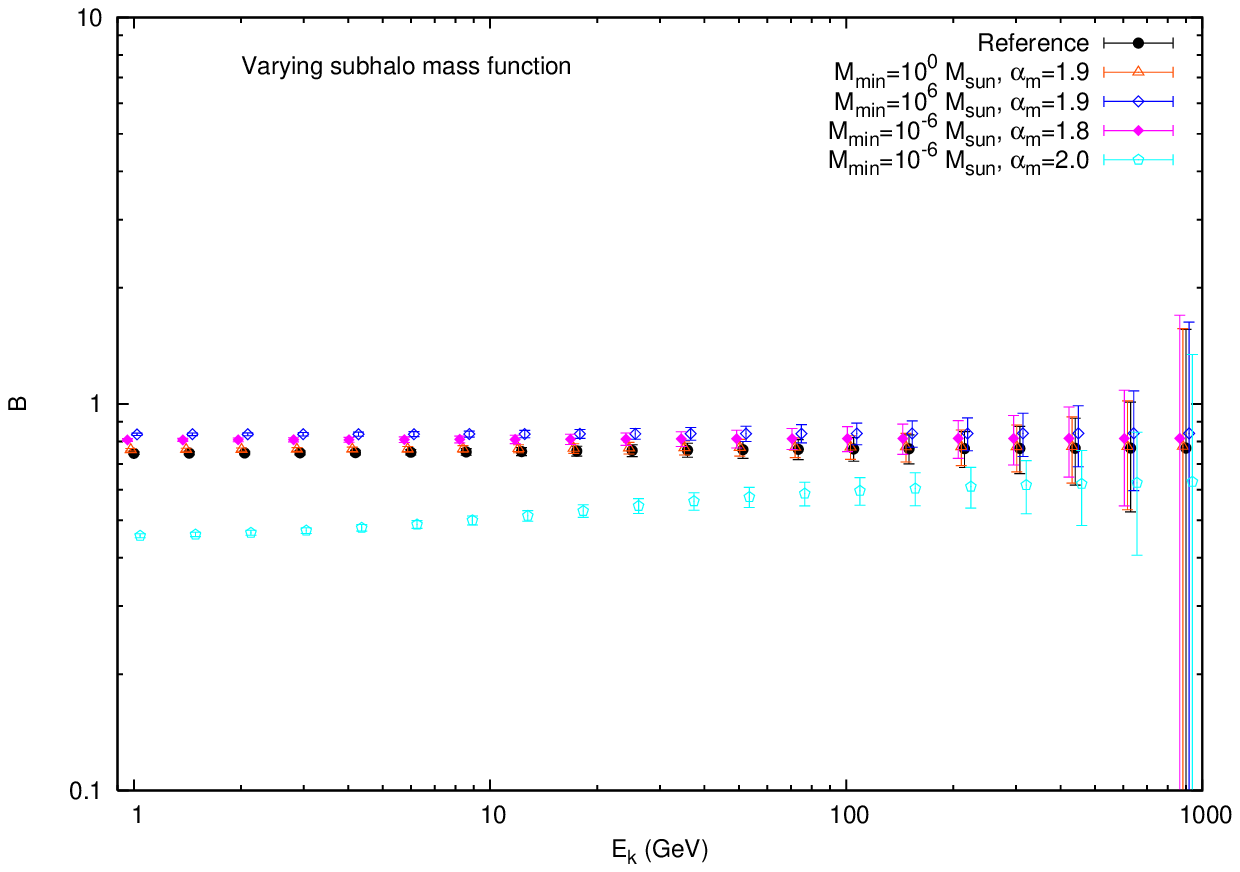}
\includegraphics[width=0.45\columnwidth]{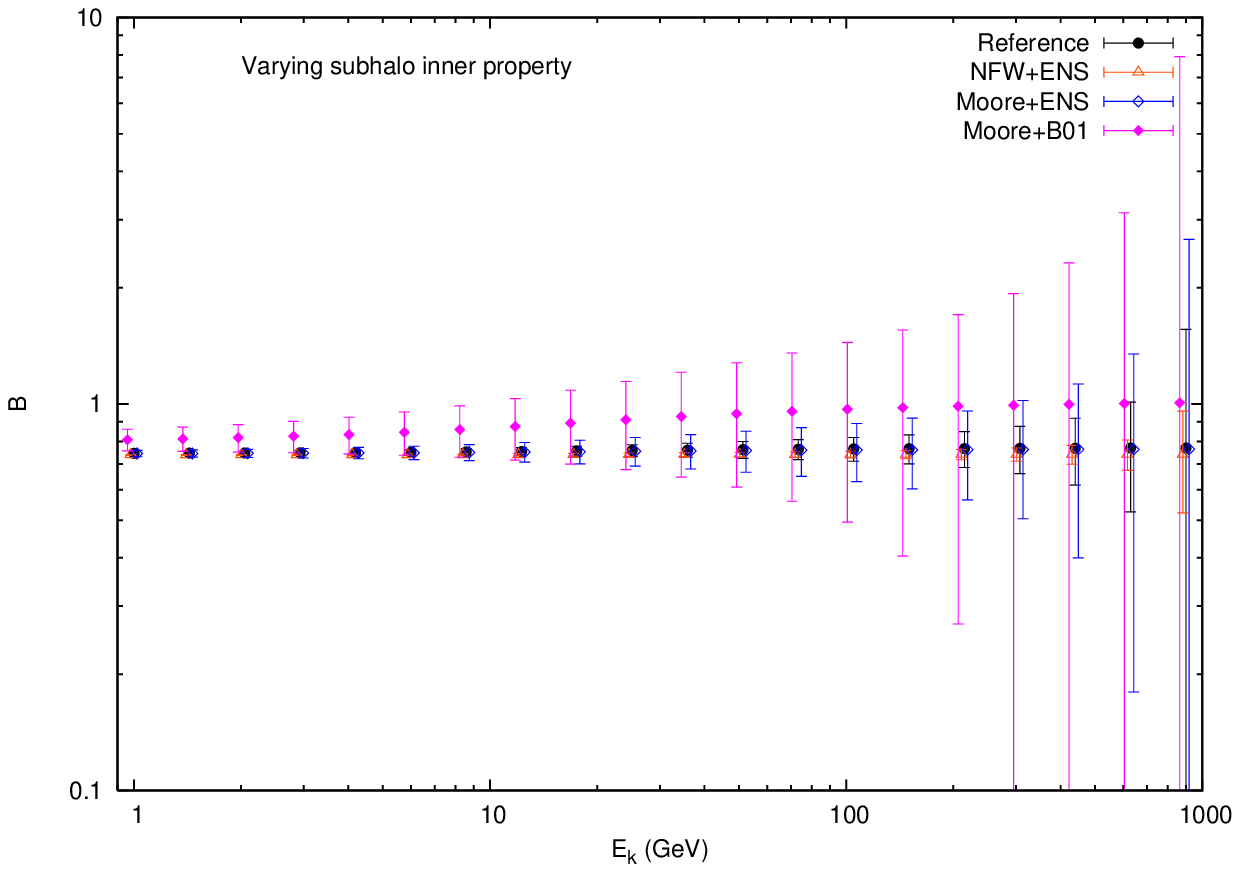}
\includegraphics[width=0.45\columnwidth]{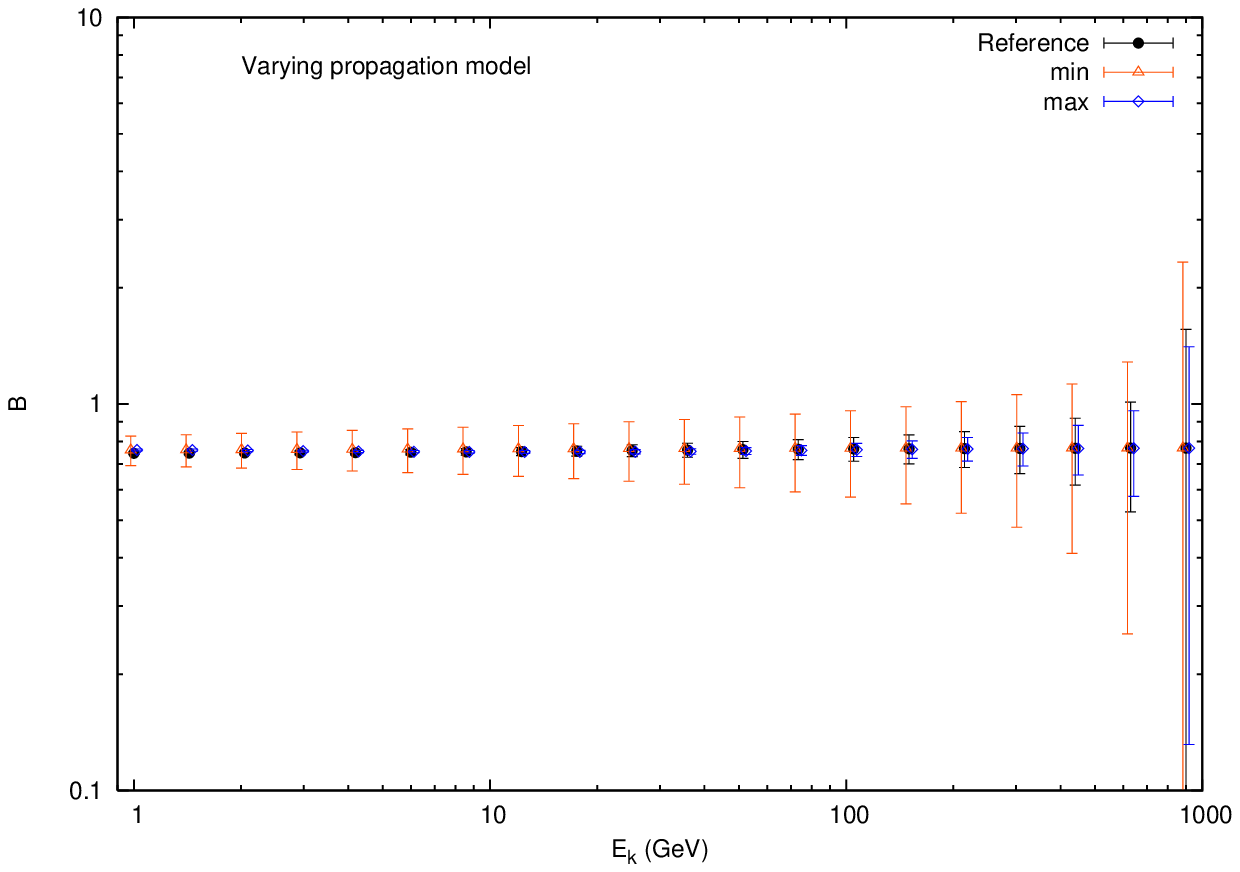}
\includegraphics[width=0.45\columnwidth]{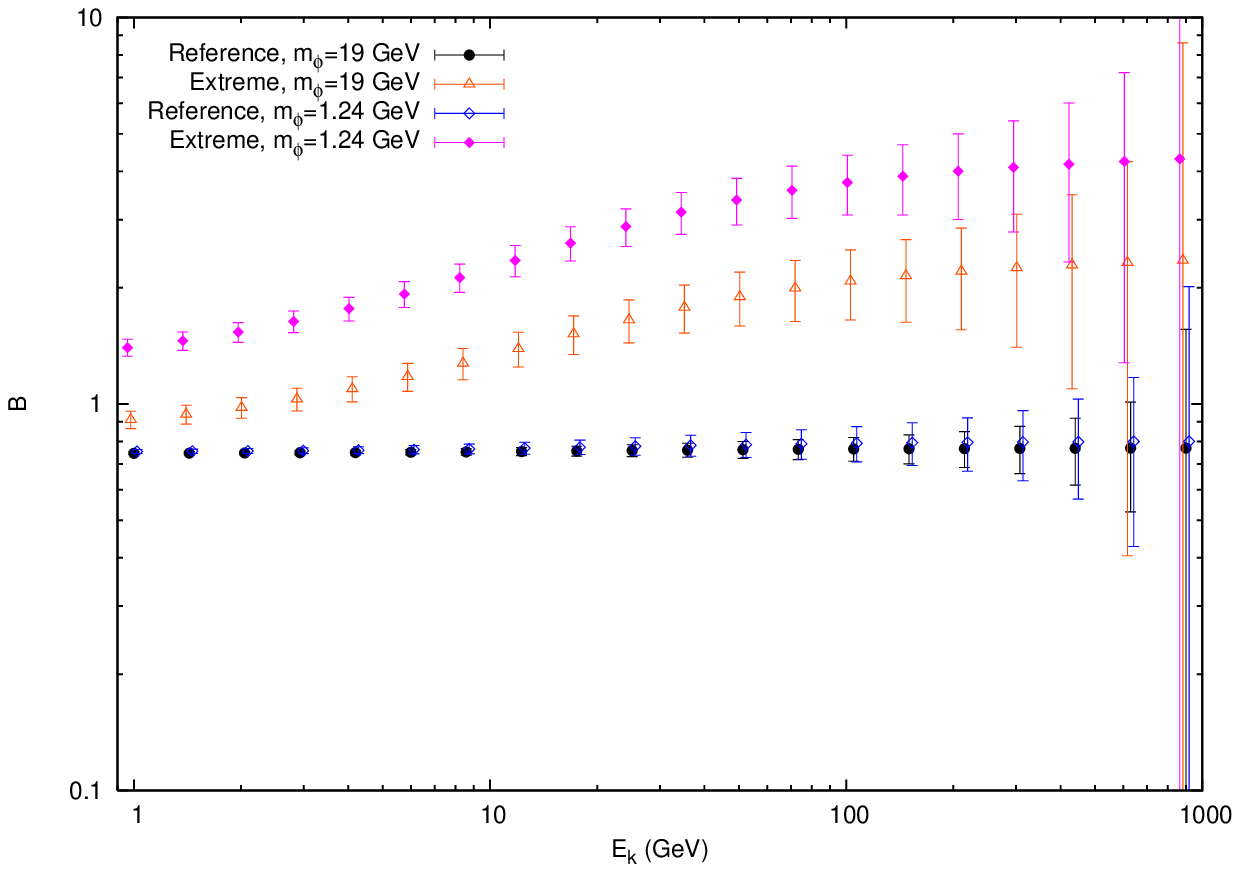}
\caption{The same as Fig. \ref{fig:pos1} but for the moderately resonant SE
enhancement case with $m_{\phi}=19$ or $1.24$ GeV.
}
\label{fig:pos2}
\end{center}
\end{figure}

\begin{figure}[htb]
\begin{center}
\includegraphics[width=0.45\columnwidth]{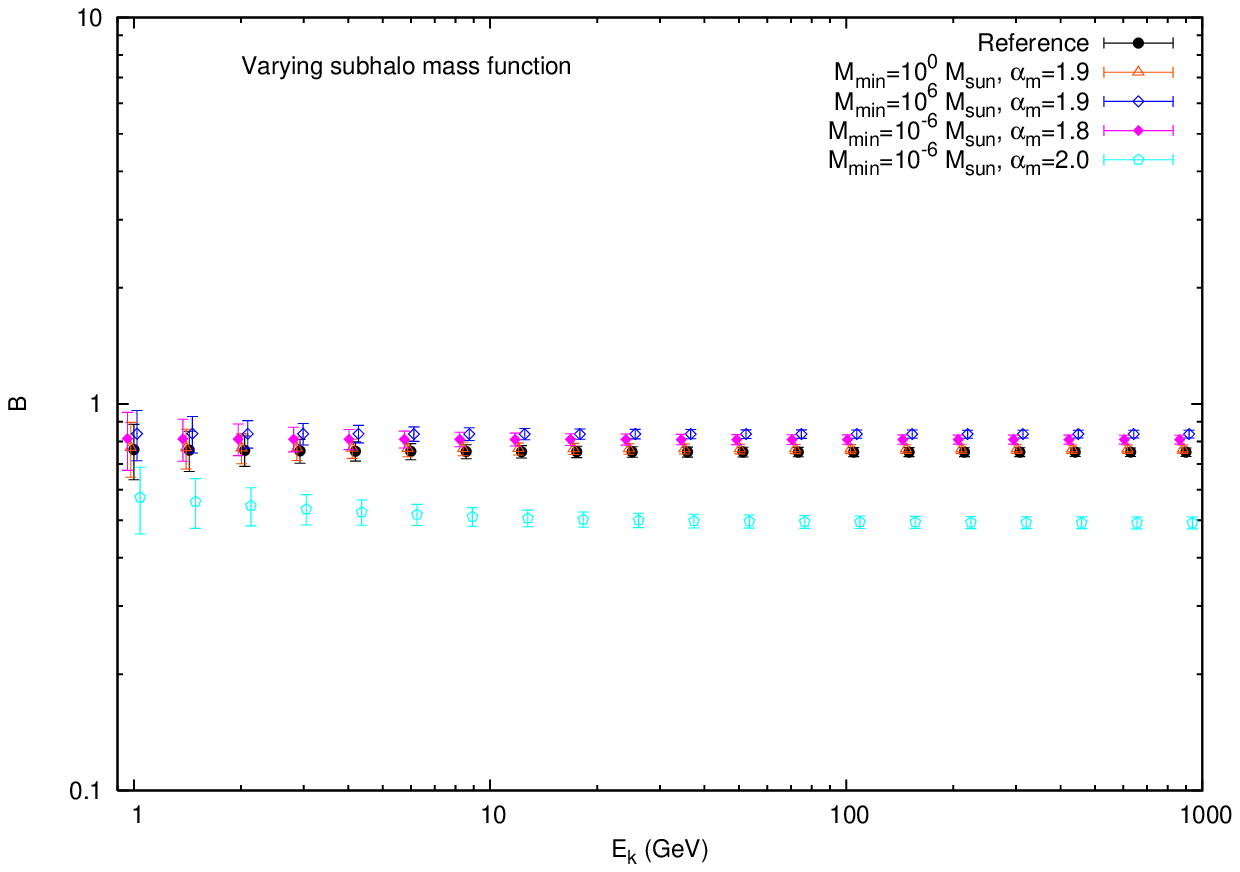}
\includegraphics[width=0.45\columnwidth]{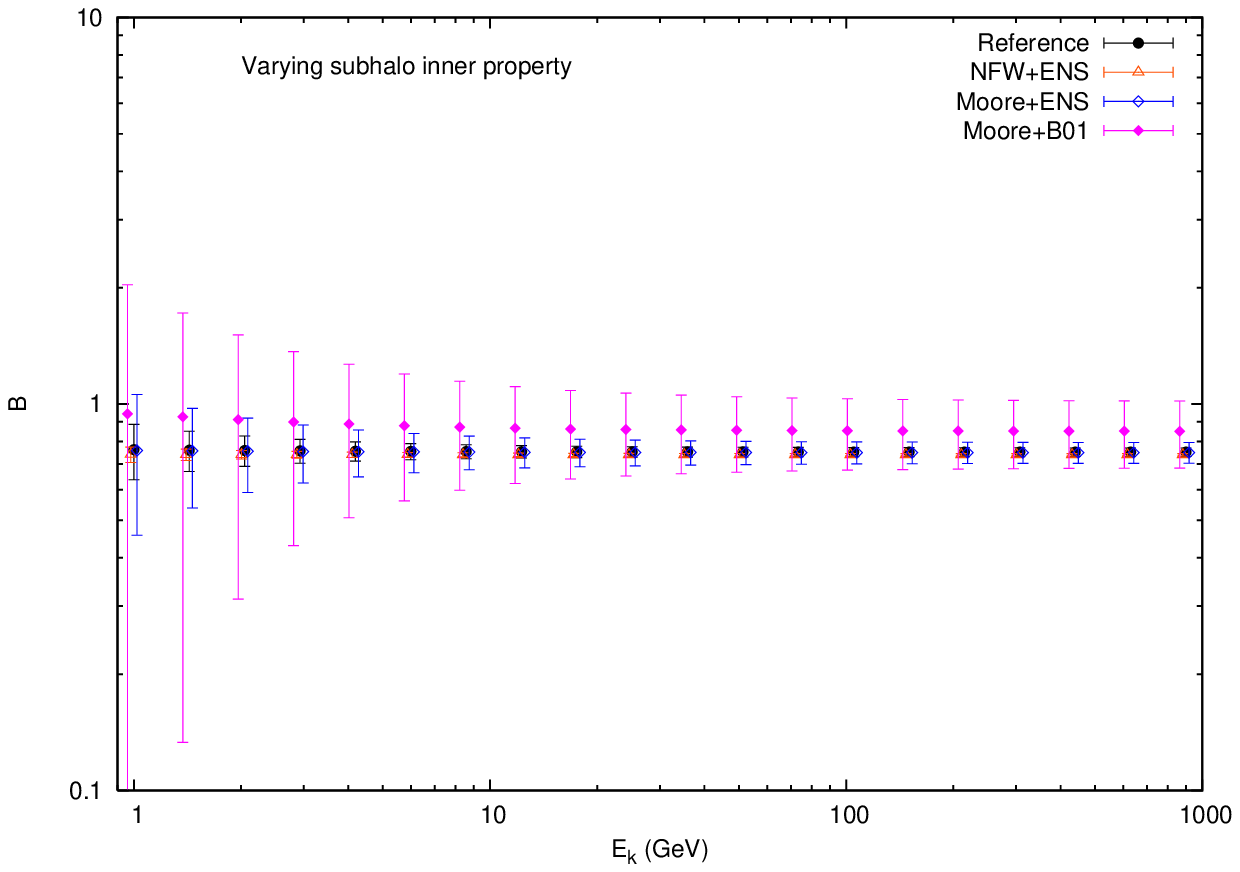}
\includegraphics[width=0.45\columnwidth]{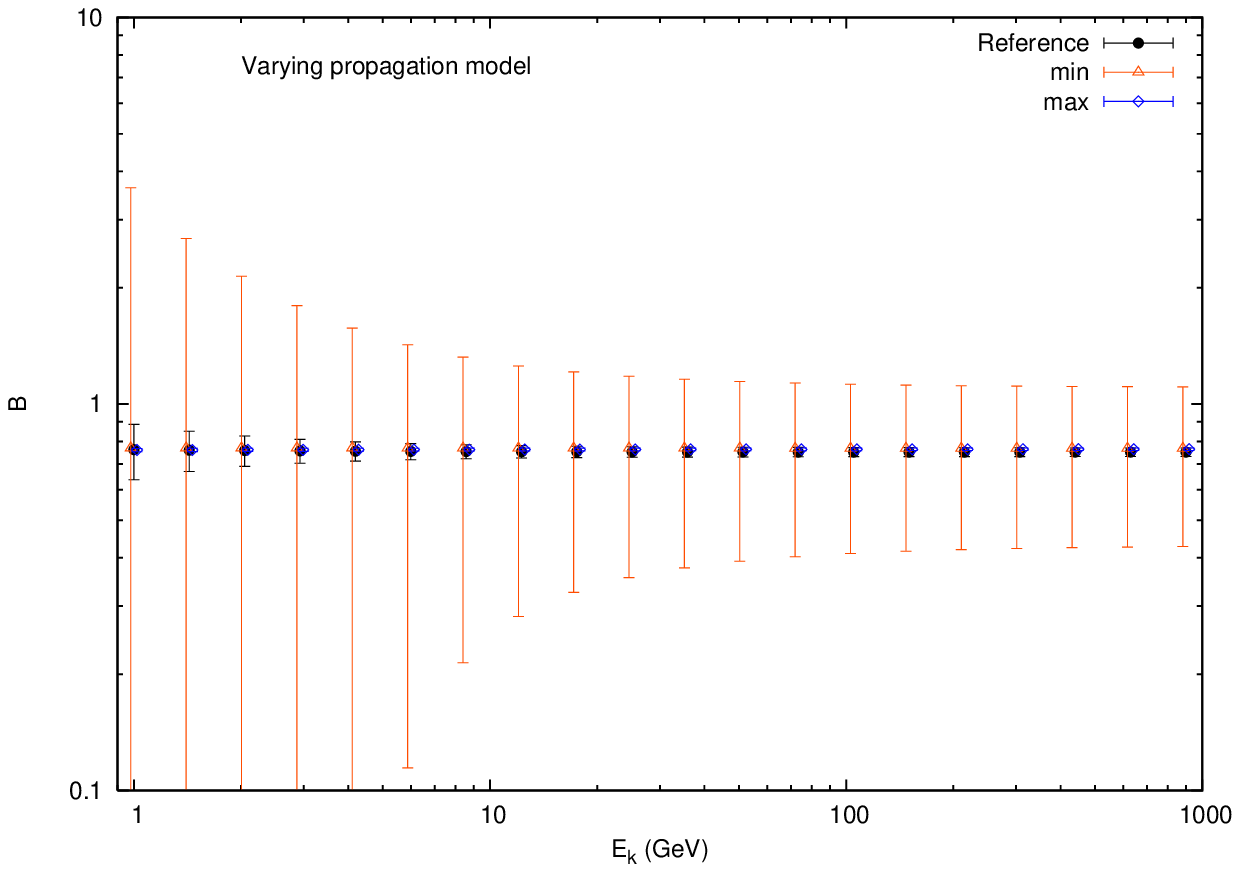}
\includegraphics[width=0.45\columnwidth]{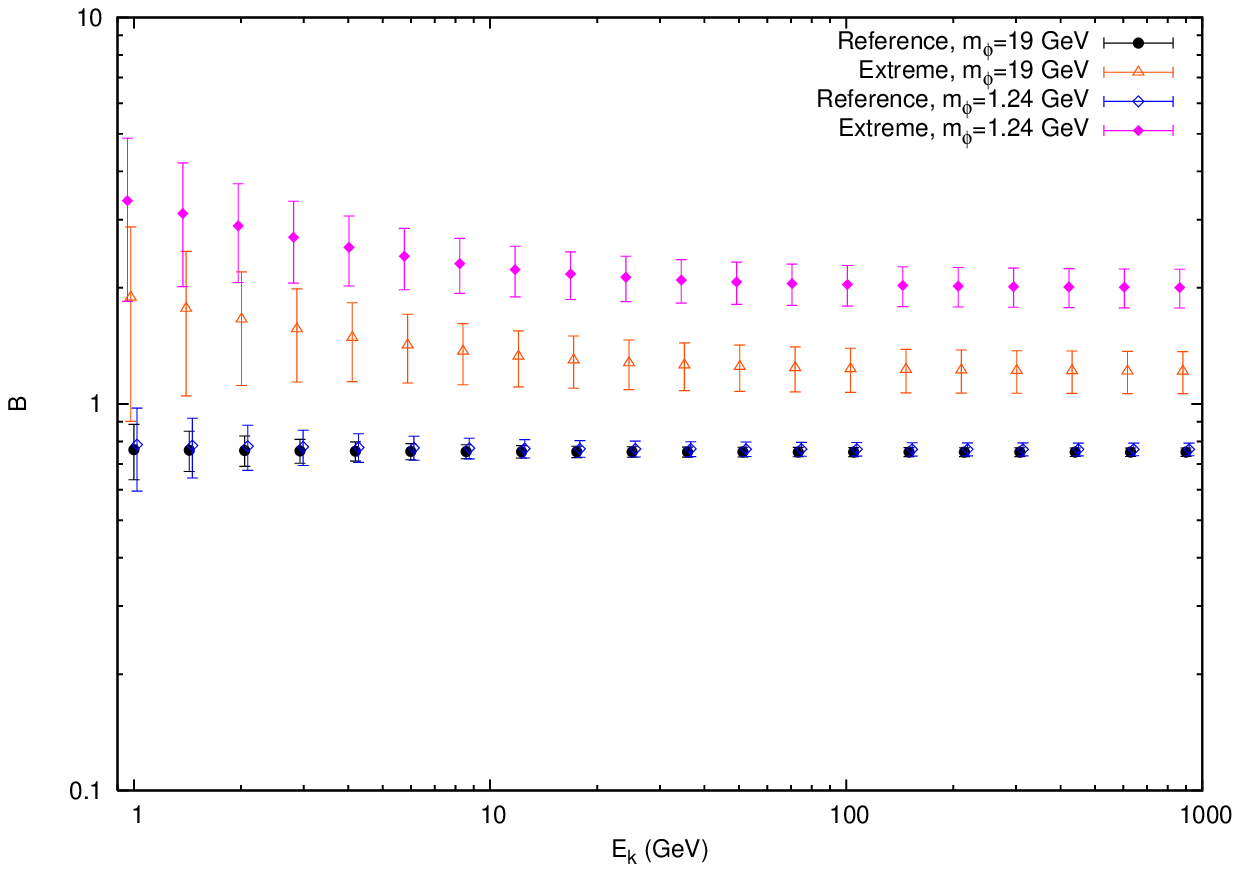}
\caption{The same as Fig. \ref{fig:pos2} but for antiprotons.
}
\label{fig:pba2}
\end{center}
\end{figure}

If the smooth halo lies out of the saturation region of SE, we may
expect stronger boost effects from substructures. For the
moderately resonant case with $m_{\phi}=19$ GeV (or $1.24$ GeV),
the average SE enhancement factor for the saturation region is
$\sim 2$ (or $4$) times larger than that of the smooth MW halo.
Similar with the previous discussion of $m_{\phi}=1$ GeV case, we
will expect $\sim 2-4$ times larger boost effect than the case
without SE. Detailed calculation gives exactly the expected
results, as shown in Figs. \ref{fig:pos2} and \ref{fig:pba2}. The
detailed properties and the parameter-dependences of $B$ and
$\sigma_{\rm B}$ are similar with the previous subsection. We only
point out that for the extreme case (B01 + Moore + $\alpha_{\rm
m}=2.0$) the boost factor is still less than, e.g. $\sim 5$.

\subsection{Strongly resonant case}

\begin{figure}[htb]
\begin{center}
\includegraphics[width=0.45\columnwidth]{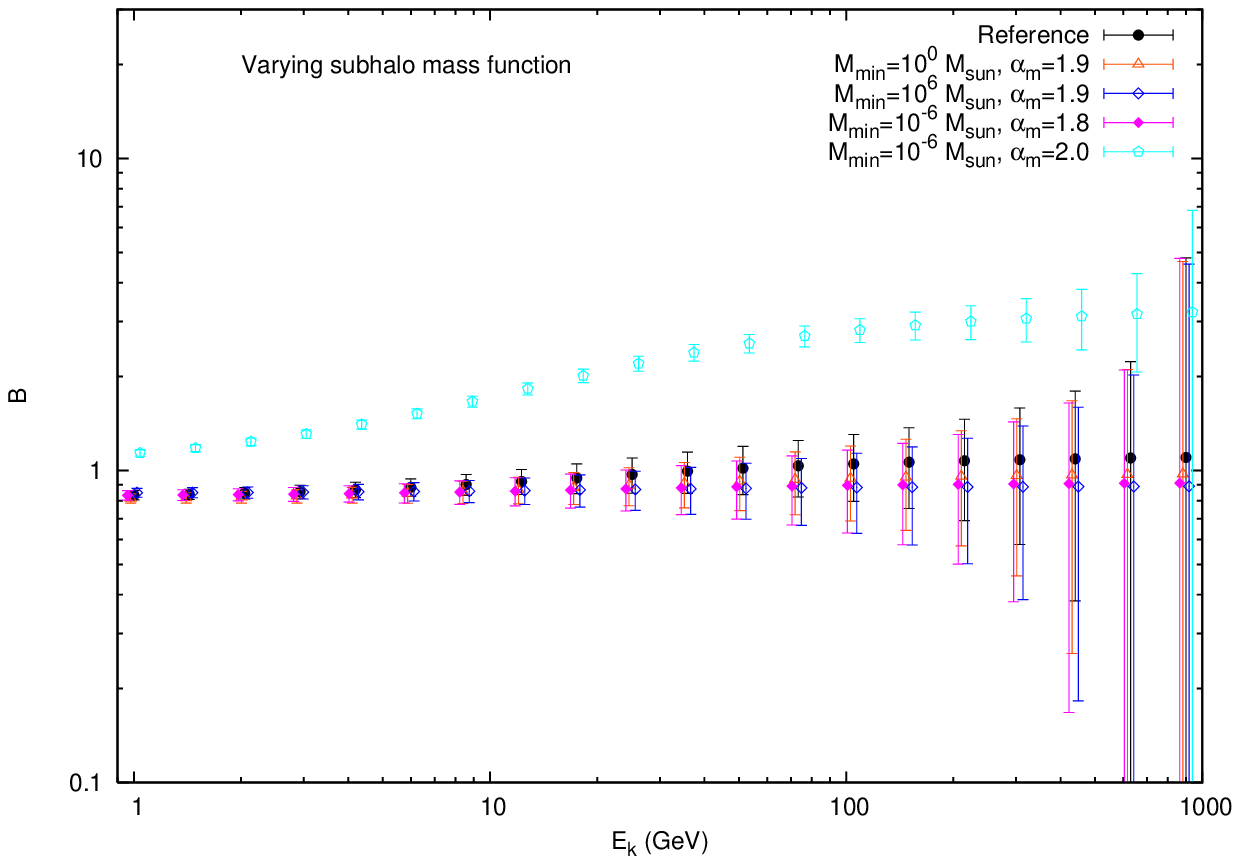}
\includegraphics[width=0.45\columnwidth]{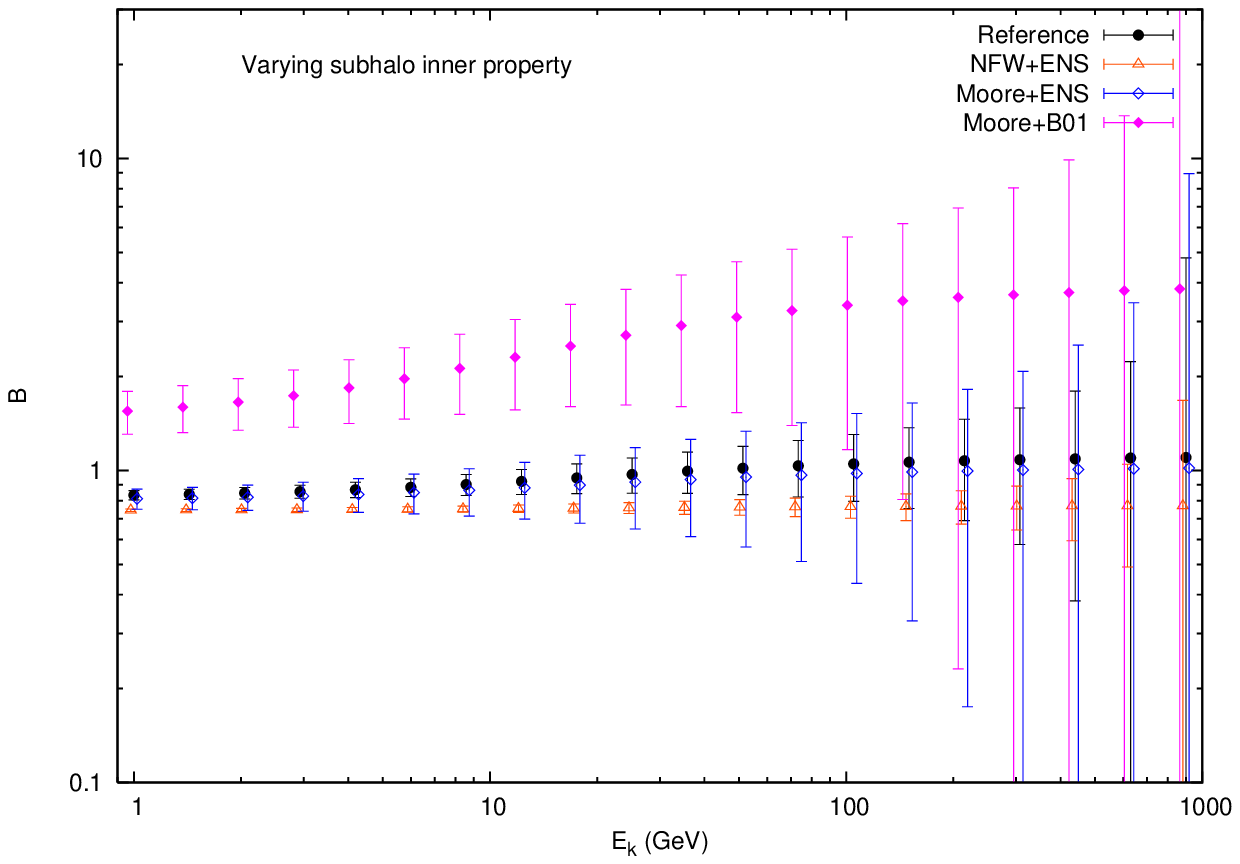}
\includegraphics[width=0.45\columnwidth]{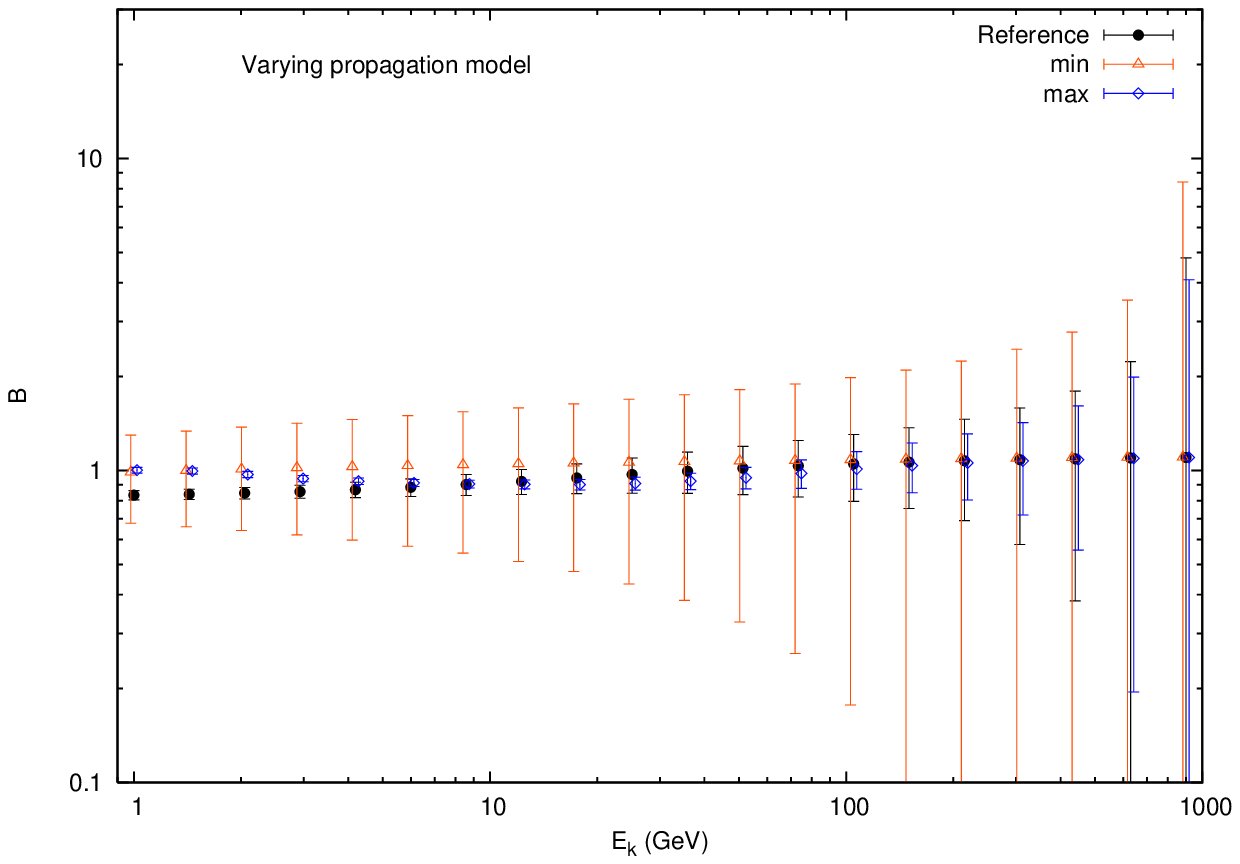}
\includegraphics[width=0.45\columnwidth]{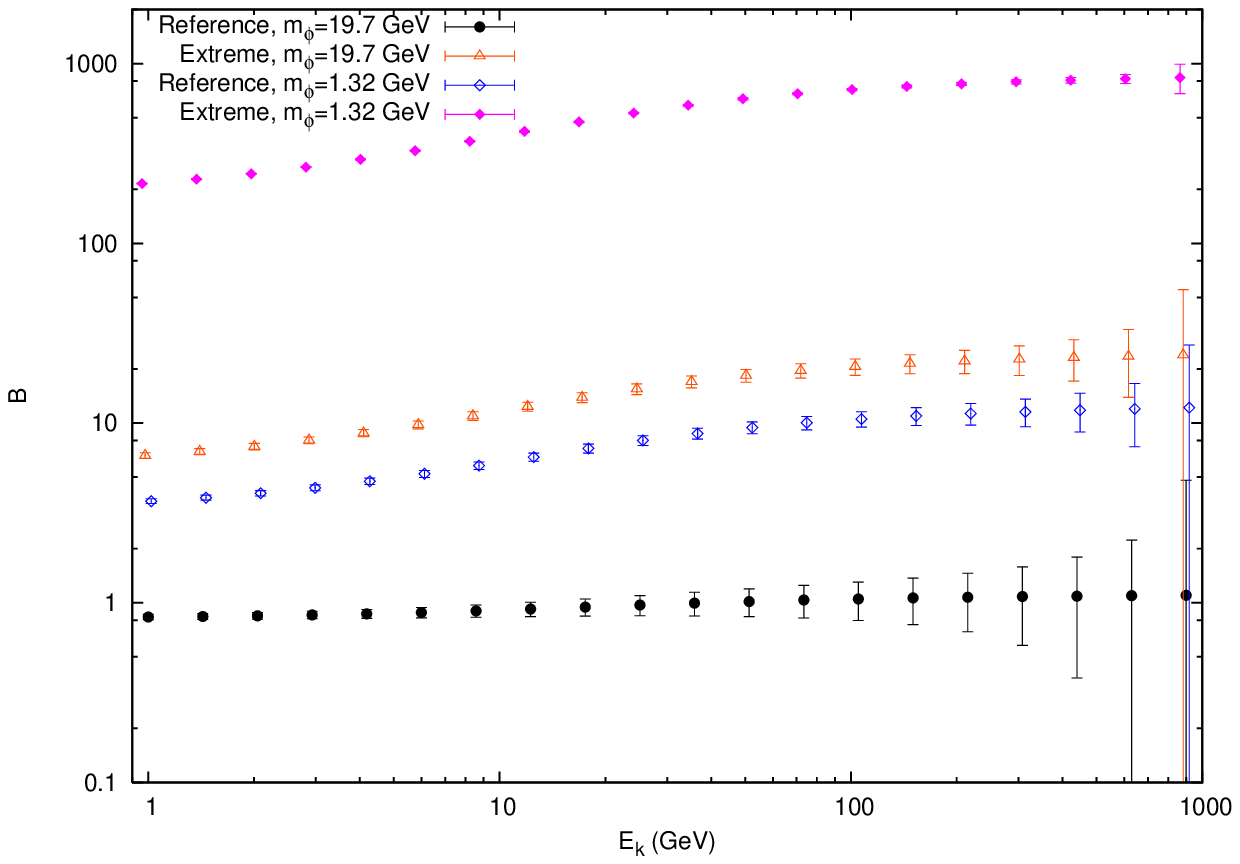}
\caption{The same as Fig. \ref{fig:pos1} but for strongly resonant case
$m_{\phi}=19.7$ or $1.32$ GeV.
}
\label{fig:pos3}
\end{center}
\end{figure}

\begin{figure}[htb]
\begin{center}
\includegraphics[width=0.45\columnwidth]{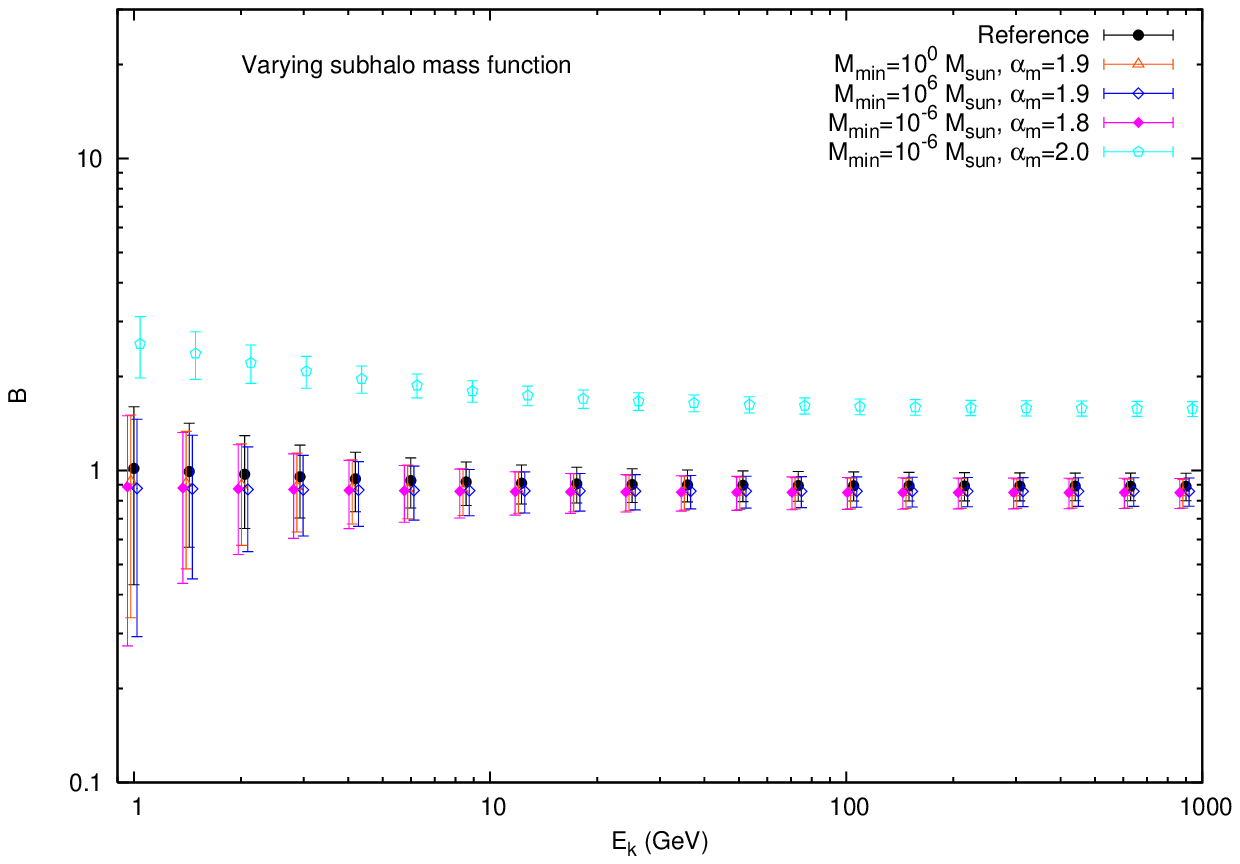}
\includegraphics[width=0.45\columnwidth]{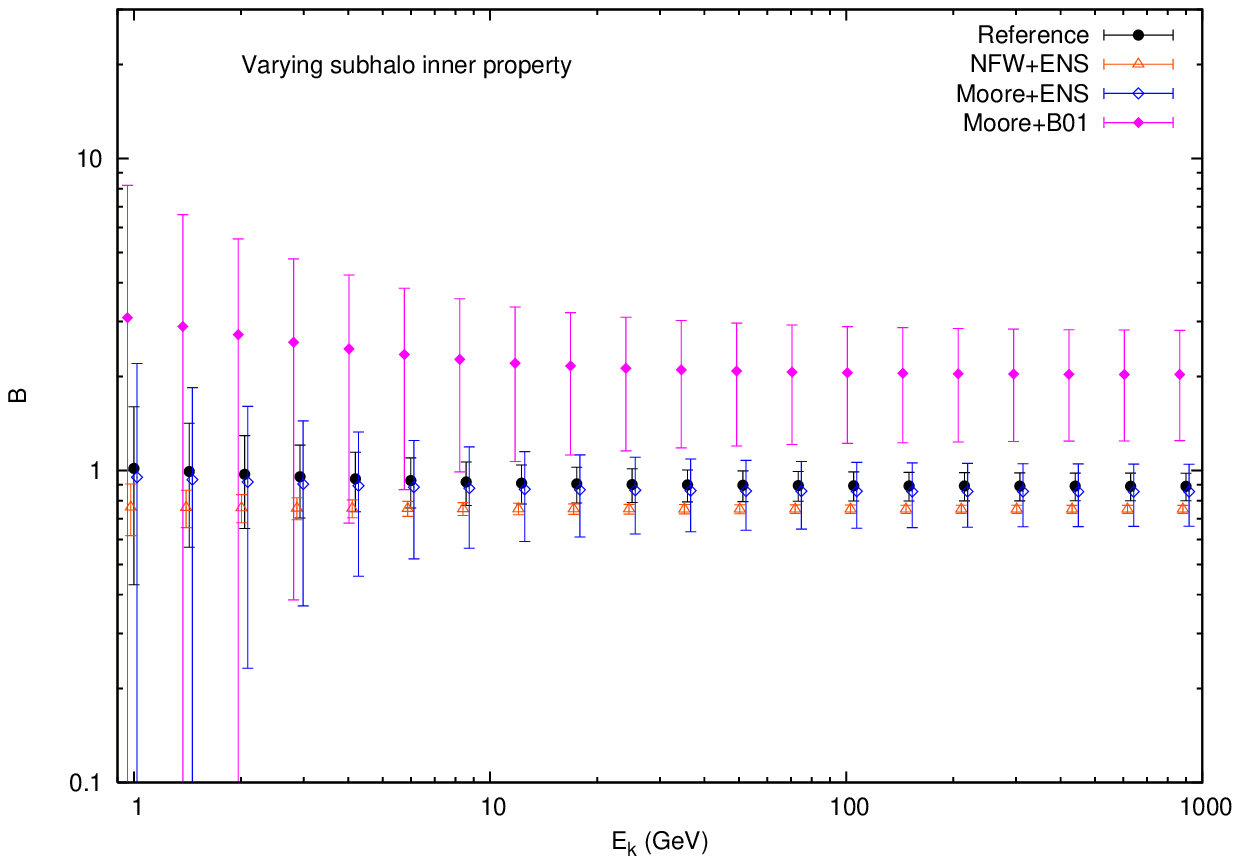}
\includegraphics[width=0.45\columnwidth]{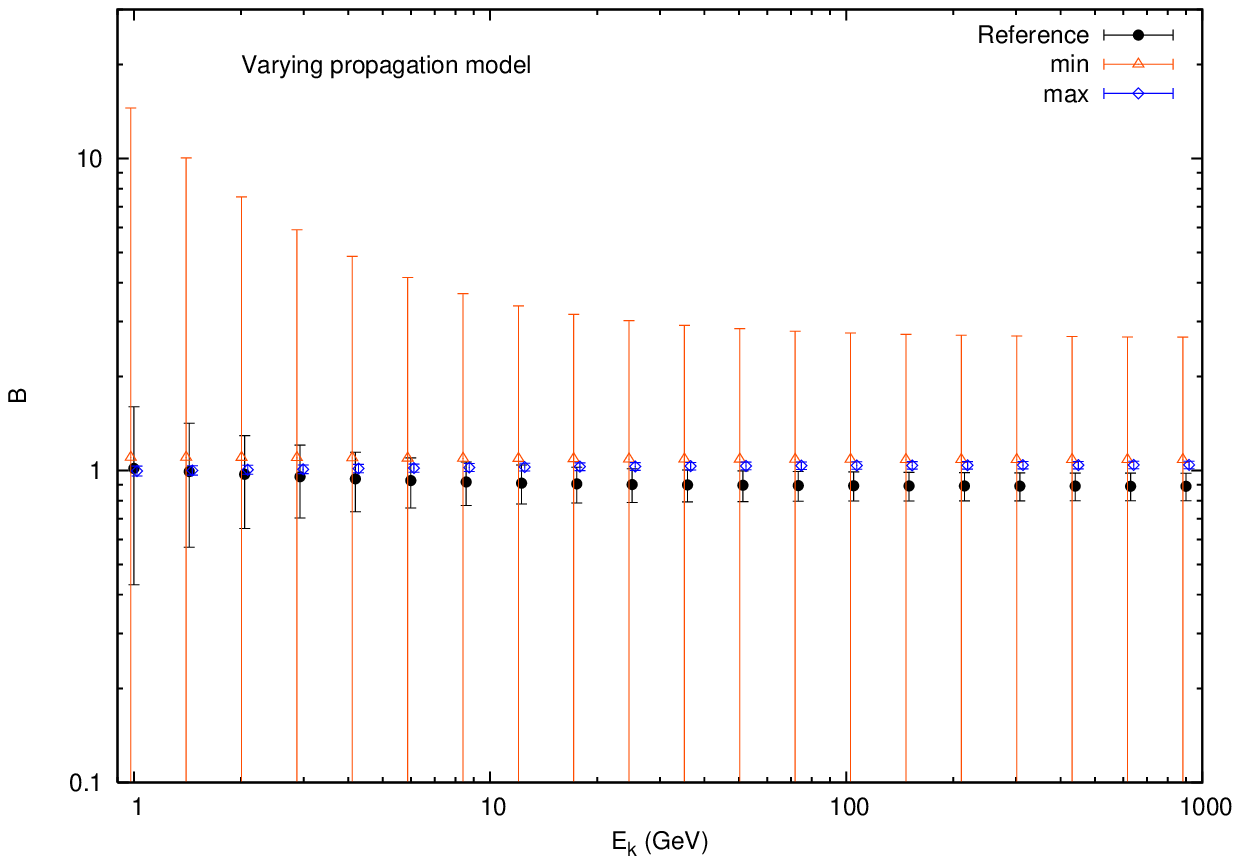}
\includegraphics[width=0.45\columnwidth]{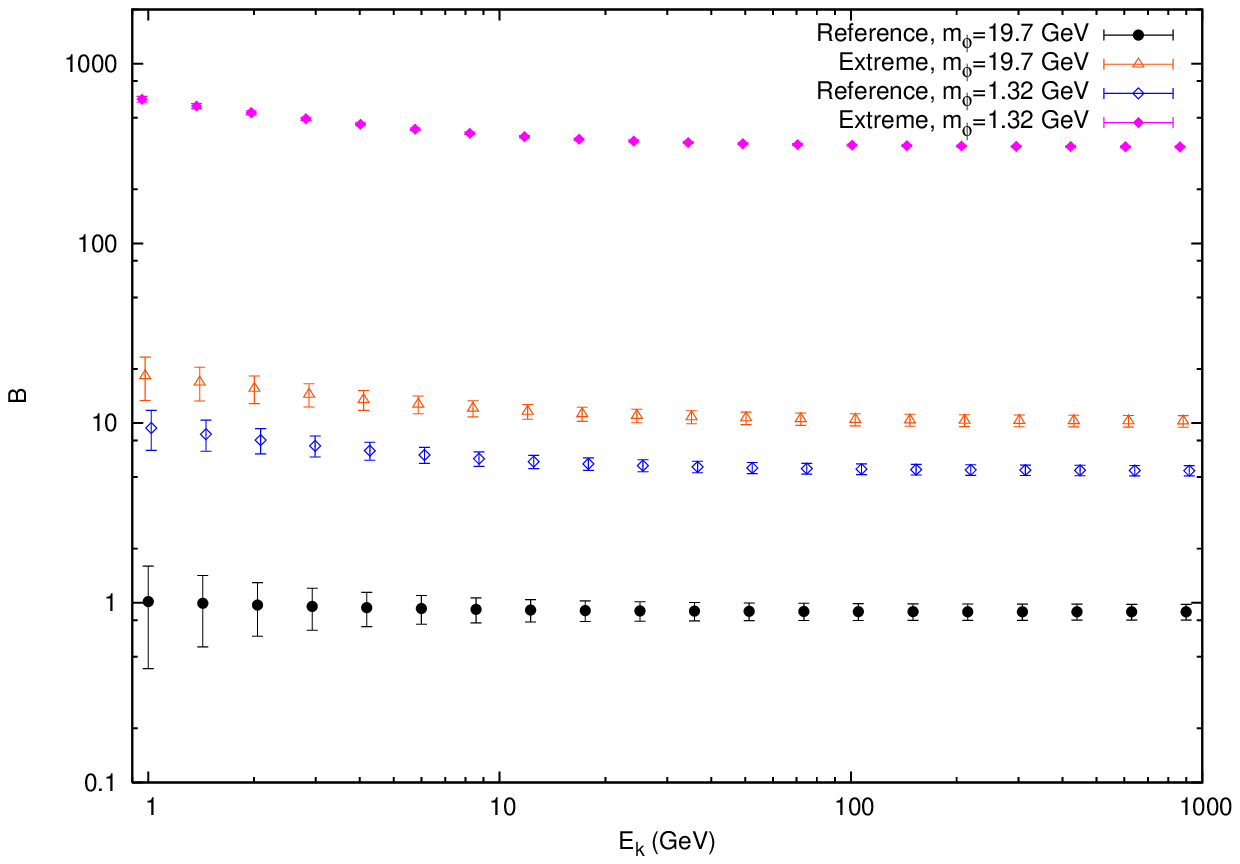}
\caption{The same as Fig. \ref{fig:pos3} but for antiprotons.
}
\label{fig:pba3}
\end{center}
\end{figure}

The results for the strongly resonant case with $m_{\phi}=19.7$ or
$1.32$ GeV are shown in Figs. \ref{fig:pos3} and \ref{fig:pba3}.
According to Fig. \ref{fig:sommer} we know that the saturation
value of the SE enhancement factor is about $20$ and $10^3$ times
larger than that of the smooth MW halo for $m_{\phi}=19.7$ and
$1.32$ GeV respectively. For the reference model configuration, we
still find almost no boost effect for $m_{\phi}=19.7$ GeV. While
for $m_{\phi}=1.32$ GeV case the final boost factor from
substructures can be as large as $\sim 10$ from the bottom-right
panel of Figs. \ref{fig:pos3} and \ref{fig:pba3}.

For the extreme case of the combinations of all maximal
configurations, the $maximal$ boost factor is found to be about
$\sim 20$ and $\sim 10^3$ for $m_{\phi}=19.7$ and $1.32$ GeV
respectively, see the bottom-right panel of Figs. \ref{fig:pos3}
and \ref{fig:pba3}. We also notice the variance of
the extreme case if very small. This is because the main
contribution comes from the very light microhalos, which have
large number density. We can conclude that if the SE is finely
tuned to be very close to the resonance peak and at the same time
taking very cuspy DM density profile and subhalo mass function
giving huge number of microhalos, the boost effect from
substructures can be remarkable. However, we think this case is
not favorable. On one hand the fine tuning of the SE parameter 
(e.g., $\phi$ in this work) is needed\footnote{If we define the degree
of fine tuning as $\eta=\Delta m_\phi/m_\phi$, it is found $\eta\lesssim
1\%$ is needed to get observable boost factors.}\footnote{Note also that
for a very small value of $m_{\phi}$, e.g. $m_{\phi}/m_{\chi}\lesssim
10^{-5}$, the saturation velocity can be of the order $m_{\phi}/m_{\chi}$
\cite{Kuhlen:2009is}, which may lead to a non-negligible boost factor
even for the non-resonant case. However, this can be regarded as another
kind of fine tuning.}. On the other hand 
the very cuspy subhalo density profile, like the Moore profile, is not 
favored by the recent high precision simulations 
\cite{Navarro:2003ew,Diemand:2005vz}. 
In some of these simulations the density profile is even shallower 
than the NFW profile \cite{Navarro:2003ew}. We may expect in this case 
the boost factor due to DM clumps should be negligible no matter the 
SE exists or not.

\section{Summary and discussion}

It has been shown that the DM clumpiness in the MW halo do not tend to
enhance the local observed fluxes of charged antiparticles such as
antiprotons and positrons, according to the N-body simulations of DM
structure formation (Paper I). In this work we re-investigate this
problem taking into account the additional boost effects in DM subhaloes
from the SE.

We find that generally the SE, if exists, has the same enhancement
effect on the smooth component and the substructures due to the
saturation behavior, except for finely tuning the SE parameter to 
some extreme cases when the saturation velocity is much smaller than
the velocity dispersion of the smooth halo. Therefore for most cases, 
the conclusions of the boost factors from DM clumpiness are similar
with Paper I. For the moderately resonant case like $m_{\phi}=19$
or $1.24$ GeV in Fig. \ref{fig:sommer}, the SE enhancement factor
for substructures with respect to the smooth halo is also small.
The total boost factors from DM subhaloes are also negligible for
the general cases of the subhalo model (mass function slope
$\alpha_{\rm m}\lesssim 2.0$, B01 concentration model, NFW or
Moore inner density profile) given in simulations, as shown in the
upper panels of Figs. \ref{fig:pos2} and \ref{fig:pba2}. Even for
the combinations of several extreme aspects of subhalo
configurations, e.g., B01 concentration + Moore inner profile +
$\alpha_{\rm m}=2.0$, the boost factor is as large as several.

For the strongly resonant case, e.g. $m_{\phi}=19.7$ or $1.32$
GeV, the additional enhancement factor for substructures from SE
is much larger. Even so, for the reference configuration of DM
distribution which is favored by simulations, we find the total
boost factors will be less than $\sim 10$. The largest boost
factors for the very extreme case of subhalo configurations (B01
concentration + Moore inner profile + $\alpha_{\rm m}=2.0$) can be
about $\sim 10$ to at most $\sim 10^3$, depending on the SE
parameter adoption.

Finally we simply address the implications of DM indirect searches
in the GC. As we have shown, in general cases the substructures do
not play a significant role for the enhancement of DM annihilation
in the solar neighborhood, as required by the PAMELA and ATIC
data. Therefore, if the DM interpretation of the observational
data is correct, the enhancement to the annihilation rate of local
DM seems also hold for the GC if there are no other spatial
dependent enhancement mechanisms. Thus the photon emission from
the GC will be a powerful tool to cross check the self-consistency
of the theory and constrain the DM density in the GC. Alternatively,
since the radio and $\gamma$-ray emission from the GC in DM
annihilation scenario is strongly constrained by the observational
data \cite{Bertone:2008xr,Zhang:2008tb,Bergstrom:2008ag,Cirelli:2009vg},
it may imply an appeal for a non-negligible contribution to the local 
DM annihilation from the clumpiness, although the DM substructures
alone can not fully account for the needed large cross section.

\appendix

\section{Solutions of the propagation equations of positrons and
antiprotons}

\subsection{Antiprotons}

It has been shown that for the propagation of antiprotons neglecting
the continuous energy losses and reacceleration can provide a good
enough approach, especially for energies higher than several GeV
\cite{Maurin:2006hy}. We will also adopt this approximation here.
Therefore the relevant processes include the diffusion, convection
and the catastrophic losses --- inelastic scattering and annihilation
in interactions. The propagation equation is
\begin{equation}
-D\Delta N+V_{\rm c}\frac{\partial N}{\partial z}+2h\Gamma_{\rm tot}
\delta(z)N=q({\bf x},E),
\end{equation}
where $\Gamma_{\rm tot}=\sum_{i=H,He}n_i\,\sigma_i^{\bar{p}}\,v$ is
the destruction rate of antiprotons in the thin gas disk with half
height $h\approx 0.1$ kpc \cite{Maurin:2006hy}, $q({\bf x},E)$ is the
source function. The propagator for a point source located at
${\bf x}_{\rm S}$, expressed in cylindrical coordinates $(r,z)$ (symmtric
in $\theta$) is \cite{Maurin:2006hy}
\begin{equation}
{\cal G}^{\bar{p}}_{\odot}(r,z,E)=\frac{\exp(-k_vz)}{2\pi DL}\times\sum_{n=0}
^{\infty}c_n^{-1}K_0\left(r\sqrt{k_n^2+k_v^2}\right)\sin(k_nL)
\sin[k_n(L-z)],
\label{proppbar}
\end{equation}
where $r$ and $z$ are the radial distance and vertical height of the
source, $K_0(x)$ is the modified Bessel function of the second type,
$k_v=V_c/2D$, and $k_n$ is the solution of the equation
$2k_n\cos(k_nL)=-(2h\Gamma_{\rm tot}/D+2k_v)\sin(k_nL)$, and
$c_n=1-\frac{\sin(k_nL)\cos(k_nL)}{k_nL}$. For any source function
$q(r,z,\theta;E)$, the local observed flux is
\begin{equation}
\Phi_{\odot}^{\bar{p}}(E)=\frac{v}{4\pi}\times 2\int_0^L{\rm d}z
\int_0^{R_{\rm max}}r{\rm d}r{\cal G}^{\bar{p}}_{\odot}(r,z,E)
\int_0^{2\pi}{\rm d}\theta q(r,z,\theta;E).
\label{fluxpbar}
\end{equation}

\subsection{Positrons}

For positrons the case is some different from antiprotons. The dominant
process in the propagation of positrons is energy loss due to synchrotron
and inverse Compton scattering for energies higher than $\sim$GeV. In
this paper we will neglect the convection and reacceleration of positrons,
which is shown to be of little effect for $E\gtrsim 10$ GeV (also the
interested energy range here) \cite{Delahaye:2008ua}. Then the propagation
equation is
\begin{equation}
-D\Delta N + \frac{\partial}{\partial E}\left(\frac{{\rm d}E}{{\rm d}t}
N\right)=q({\bf x},E),
\end{equation}
in which the second term in the left hand side represents the energy
losses. The energy loss rate of positrons due to synchrotron and inverse
Compton scattering in the MW can be adopted as ${\rm d}E/{\rm d}t=
-\epsilon^2/\tau_{\rm E}$, with $\epsilon=E/1{\rm\ GeV}$ and $\tau_{\rm E}
\approx 10^{16}$ s \cite{Baltz:1998xv}. We directly write down the
propagator for a point source located at $(r,z)$ from the solar location
with monochromatic injection energy
$E_{\rm S}$ \cite{Lavalle:2006vb,Lavalle:1900wn}
\begin{equation}
{\cal G}_{\odot}^{e^+}(r,z,E\leftarrow E_{\rm S})=\frac{\tau_{\rm E}}
{E\epsilon}\times \hat{\cal G}_{\odot}(r,z,\hat{\tau}),
\label{propposi}
\end{equation}
in which we define a pseudo time $\hat{\tau}$ as
\begin{equation}
\hat{\tau}=\tau_{\rm E}\frac{\epsilon^{\delta-1}-
\epsilon_{\rm S}^{\delta-1}}{1-\delta}.
\end{equation}
$\hat{\cal G}_{\odot}(r,z,\hat{\tau})$ is the Green's function for
the re-arranged diffusion equation with respect to the pseudo time
$\hat{\tau}$
\begin{equation}
\hat{\cal G}_{\odot}(r,z,\hat{\tau})=\frac{\theta(\hat{\tau})}
{4\pi D_0\hat{\tau}}\exp\left(-\frac{r^2}{4D_0\hat{\tau}}\right)
\times {\cal G}^{\rm 1D}(z,\hat{\tau}).
\end{equation}
The effect of boundaries along $z=\pm L$ appears in ${\cal G}^{\rm 1D}$
only. Following Ref. \cite{Lavalle:2006vb} we use two distinct regimes to
approach ${\cal G}^{\rm 1D}$:
\begin{itemize}

\item for $\zeta\equiv L^2/4D_0\hat{\tau}\gg 1$ (the extension of electron
sphere $\lambda\equiv\sqrt{4D_0\hat{\tau}}$ is small)
\begin{equation}
{\cal G}^{\rm 1D}(z,\hat{\tau})=\sum_{n=-\infty}^{\infty}(-1)^n
\frac{\theta(\hat{\tau})}{\sqrt{4\pi D_0\hat{\tau}}}\exp\left(
-\frac{z_n^2}{4D_0 \hat{\tau}}\right),
\end{equation}
where $z_n=2Ln+(-1)^nz$;

\item otherwise
\begin{equation}
{\cal G}^{\rm 1D}(z,\hat{\tau})=\frac{1}{L}\sum_{n=1}^{\infty}
\left[\exp(-D_0k_n^2\hat{\tau})\phi_n(0)\phi_n(z)+
\exp(-D_0k_n'^2\hat{\tau})\phi_n'(0)\phi_n'(z)\right],
\end{equation}
where
\begin{eqnarray}
\phi_n(z)&=&\sin[k_n(L-|z|)];\ \ k_n=(n-1/2)\pi/L,\\
\phi_n'(z)&=&\sin[k_n'(L-z)];\ \ k_n'=n\pi/L.
\end{eqnarray}

\end{itemize}
For any source function $q(r,z,\theta;E_{\rm S})$ the local observed
flux of positrons can be written as
\begin{equation}
\Phi_{\odot}^{e^+}=\frac{v}{4\pi}\times 2\int_0^L{\rm d}z
\int_0^{R_{\rm max}}r{\rm d}r\int_E^{\infty}{\rm d}E_{\rm S}
{\cal G}^{e^+}_{\odot}(r,z,E\leftarrow E_{\rm S})
\int_0^{2\pi}{\rm d}\theta q(r,z,\theta;E_{\rm S}).
\label{fluxposi}
\end{equation}

\subsection{Propagation parameters}

The propagation parameters are determined by fitting the local
observed CR spectra such as the B/C ratio and unstable-stable
isotope ratio. In Ref. \cite{Maurin:2001sj} the authors showed
that there are strong degeneracies between parameters to recover
the local measured B/C ratio. In this work we adopt three typical
settings of parameters which are consistent with the B/C data, but
give very different primary anti-particle fluxes from DM
annihilation \cite{Donato:2003xg}. These parameters, labelled as
``max'', ``med'' and ``min'' according to the primary antiparticle
fluxes, are gathered in Table \ref{table:prop}.

\begin{table}[!htb]
\begin{center}
{\begin{tabular}{c c c c c}
\hline
\hline
&  $\delta$  & $D_0$ (kpc$^{2}$~Myr$^{-1}$) & $L$ (kpc)   &
$V_{\rm c}$ (km~s$^{-1})$  \\
\hline
{\rm max} &  0.46  & 0.0765 & 15 & 5.0    \\
{\rm med} &  0.70  & 0.0112 & 4  & 12.0   \\
{\rm min} &  0.85  & 0.0016 & 1  & 13.5 \\
\hline
\end{tabular}}
\caption{
Propagation parameters which are compatible with B/C analysis while giving
the maximal, median and minimal anti-particle DM fluxes.
\label{table:prop}}
\end{center}
\end{table}

\section{Boost factors}

In this section, we calculate the fluxes of charged anti-particles and
the corresponding boost factors. The source function of antiprotons
and positrons from DM annihilation can be written as
\begin{equation}
q({\bf x},E)=\frac{\langle\sigma v\rangle}{2m_{\chi}^2}
\frac{{\rm d}N}{{\rm d}E}\rho^2({\bf x}),
\end{equation}
where $m_{\chi}$ is the mass of DM particle which is set to be
$\sim 1$ TeV in light of the recent observations of ATIC
\cite{Chang:2008zz} and Fermi \cite{fermi}, $\langle\sigma
v\rangle$ is the thermally averaged velocity weighted cross
section, and ${\rm d}N/{\rm d}E$ is the yield spectrum of
$\bar{p}$ or $e^+$ per pair annihilation. As discussed in Sec. II,
the cross section is velocity-dependent if the SE is taken into
account.

Then we can rewrite the fluxes of $\bar{p}$ and $e^+$, Eqs.(\ref{fluxpbar})
and (\ref{fluxposi}), for the smooth halo as
\begin{equation}
\Phi_{\rm sm}(E)=\frac{v}{4\pi}\times A \times \int {\rm d}^3{\bf x}
\rho_{\rm sm}^2({\bf x}) \tilde{\cal G}(r,z,E)\bar{S}(\sigma),
\label{fluxsm}
\end{equation}
where $A=\langle\sigma v\rangle_0/2m_{\chi}^2$, $\bar{S}(\sigma)$ is the
average SE enhancement factor, and $\tilde{\cal G}$ is the pseudo
propagator (Paper I) which absorbes the source energy
spectrum ${\rm d}N/{\rm d}E$ in the propagator of Eq.(\ref{proppbar})
or (\ref{propposi})
\begin{eqnarray}
\tilde{\cal G}^{\bar{p}}(r,z,E)&=&\frac{{\rm d}N}{{\rm d}E}\times
{\cal G}^{\bar{p}}_{\odot}(r,z,E),\\
\tilde{\cal G}^{e^+}(r,z,E)&=&\int_E^{\infty}{\rm d}E_{\rm S}
\frac{{\rm d}N}{{\rm d}E_{\rm S}}\times {\cal G}^{e^+}_{\odot}
(r,z,E\leftarrow E_{\rm S}).
\end{eqnarray}
Note that in Eq.(\ref{fluxsm}), the SE factor should be spatially
dependent due to different velocity dispersion of DM particles in
the MW. Since the charged particles are thought to come from
places not very far from us, we will use $\sigma_{\odot} \approx
150$ km s$^{-1}$ to represent the average velocity dispersion of
the smooth DM halo for simplicity.

In our discussion we simplify the source spectrum ${\rm d}N/{\rm
d}E$ as in Paper I: for $\bar{p}$ we adopt ${\rm d}N/{\rm d}E=1$
GeV$^{-1}$, while for $e^+$ we set ${\rm d}N/{\rm
d}E=\delta(E-m_{\chi})$. This adoption makes the following
discussion independent of detailed particle physics model of DM,
while the major property about the boost factor is still kept. It
will be easy to convolve any model predicted DM source spectrum on
these results.

The flux from DM subhaloes is thought to be the sum of the population
of DM point sources
\begin{equation}
\Phi_{\rm sub}=\sum_{i=1}^{N_{\rm sub}}\Phi_i=\frac{vA}{4\pi}
\sum_{i=1}^{N_{\rm sub}}\xi_i\times \tilde{\cal G}(r_i,z_i,E),
\end{equation}
where $\xi_i\equiv \int {\rm d}^3{\bf x}'\rho^2({\bf x}')\bar{S}(\sigma_i)$
is the annihilation luminosity of the $i$th subhalo. For the velocity
dispersion of subhaloes, we adopt the relation $\sigma=2.7\times
(M_{\rm sub}/10^6M_{\odot})^{1/3}$ km s$^{-1}$ \cite{Bovy:2009zs}, with
the normalization fitted from the observational results of dwarf
spheroidals in Ref. \cite{Simon:2007dq}.

Because we do not know exactly the location and mass of any
specified subhalo, there should be uncertainty of the prediction
of flux from subhaloes. The average and relative variance of the
flux due to different realization of substructure distribution are
(Paper I)
\begin{eqnarray}
\langle\Phi_{\rm sub}\rangle&=&N_{\rm sub}\times\frac{vA}{4\pi}\times
\langle\xi\rangle_M\times\langle\tilde{\cal G}\rangle_V, \label{aver_flux}\\
\frac{\sigma_{\rm sub}^2}{\langle\Phi_{\rm sub}\rangle^2}&=&
\frac{1}{N_{\rm sub}}\left(
\frac{\sigma_{\tilde{\cal G}}^2}{\langle\tilde{\cal G}\rangle_V^2}+
\frac{\sigma_{\xi}^2}{\langle\xi\rangle_M^2}+
\frac{\sigma_{\tilde{\cal G}}^2}{\langle\tilde{\cal G}\rangle_V^2}
\times\frac{\sigma_{\xi}^2}{\langle\xi\rangle_M^2}\right),\label{rela_vari}
\end{eqnarray}
where
\begin{eqnarray}
\langle\xi\rangle_M&=&\int_{M_{\rm min}}^{M_{\rm max}}{\rm d}M_{\rm sub}
\xi(M_{\rm sub})\frac{{\rm d}P_{\rm M}}{{\rm d}M_{\rm sub}},\\
\langle\tilde{\cal G}\rangle_V&=&\int_V{\rm d}^3{\bf x}
\tilde{\cal G}\frac{{\rm d}P_{\rm V}}{{\rm d}V},\\
\sigma_{\xi}^2&=&\int_{M_{\rm min}}^{M_{\rm max}}{\rm d}M_{\rm
sub} \xi^2(M_{\rm sub})\frac{{\rm d}P_{\rm M}}{{\rm d}M_{\rm
sub}}-
\langle\xi\rangle_M^2,\\
\sigma_{\tilde{\cal G}}^2&=&\int_V{\rm d}^3{\bf x} \tilde{\cal
G}^2\frac{{\rm d}P_{\rm V}}{{\rm d}V}-\langle\tilde{\cal G}
\rangle_V^2.
\end{eqnarray}

Finally the boost factor, defined as the ratio of the sum of the
smooth and substructure contributions to the smooth one without
substructures, is\footnote{Note the defination of $\Phi_{\rm sm}$
is a bit different from Paper I, where the smooth contribution
without substructures is labelled $\Phi_{\rm sm}$.}
\begin{equation}
B=\frac{\Phi_{\rm sm}+\Phi_{\rm sub}}{\Phi_{\rm sm}^0}=(1-f)^2
(1+\Phi_{\rm sub}/\Phi_{\rm sm}),
\label{boost}
\end{equation}
where $\Phi_{\rm sm}^0$ is the smooth contribution without substructures,
and $f$ is the mass fraction of subhaloes. The variance of the boost factor
is
\begin{equation}
\sigma_{\rm B}=\frac{\sigma_{\rm sub}}{\Phi_{\rm sm}^0}=
(1-f)^2\frac{\sigma_{\rm sub}}{\Phi_{\rm sm}}.
\label{variance}
\end{equation}

\acknowledgments We thank J. Lavalle for helpful discussions on
the electron/positron propagation. This work was supported in part 
by the Natural Sciences Foundation of China (Nos. 10773011, 10775001,
10635030) and by the Chinese Academy of Sciences under the grant
No. KJCX3-SYW-N2.

\end{document}